\documentclass[twocolumn]{aastex63}
\usepackage{CJKutf8}
\usepackage{amsmath}

\newcommand{\sgra}{$\rm Sgr~A^{\star}$}

\received{}
\revised{}
\accepted{}

\submitjournal{ApJ}

\shorttitle{}
\shortauthors{Zheng, Lin, Mao}

\graphicspath{{./}{figures/}}

\begin{document}
\begin{CJK*}{UTF8}{gbsn}

\title{The influence of the secular perturbation of an intermediate-mass companion: I.  Eccentricity excitation of disk stars at the Galactic center}

\author[0000-0002-7814-9185]{Xiaochen Zheng （郑晓晨）}
\affiliation{Department of Astronomy, Tsinghua University, Beijing, China}

\author{Douglas N. C. Lin (林潮)}
\affiliation{Department of Astronomy and Astrophysics, University of California, Santa Cruz, USA}
\affiliation{Institute for Advanced Studies, Tsinghua University, Beijing, China}
 
\author{Shude Mao (毛淑德)}
\affiliation{Department of Astronomy, Tsinghua University, Beijing, China}
\affiliation{National Astronomical Observatories of China, Chinese Academy of Sciences,
20A Datun Road, Beijing, China}

\begin{abstract}
There is a dense group of OB and Wolf-Rayet stars within a fraction of a parsec from the super-massive
black hole (SMBH) at the Galactic Center.
These stars appear to be coeval and relatively massive.  A subgroup of these stars orbits on the same plane.  
If they emerged with low to modest eccentricity orbits from a common gaseous disk around the central super-massive black hole, their inferred lifespan would not be sufficiently long to account for the excitation of their high orbital eccentricity through dynamical relaxation.  Here we analyze the secular perturbation on Galactic Center stars by an intermediate-mass companion (IMC) as a potential mechanism to account for these young disk stars' high eccentricity.  This IMC may be either an intermediate-mass black hole (IMBH) or
a compact cluster such as IRS-13E.  If its orbital angular momentum vector is anti-parallel to that
of the disk stars, this perturbation would be effective in exciting the eccentricity of stars with orbital precession rates which resonate with IMC's precession rate.  If it orbits around the SMBH in the same direction as the
disk stars, the eccentricity of the young stars can still be highly excited by the IMC during the depletion of their natal disk, possible associated with the launch of the Fermi bubble. In this scenario, IMC's precession rate decreases and its secular resonance sweeps through the proximity of the young stars.  We carry out numerical simulations with various inclination angles between the orbits of IMC and the disk stars and show this
secular interaction is a robust mechanism to excite the eccentricity and inclination of some disk stars.
\end{abstract}

\keywords{methods: numerical --- stars: distances --- stars: early-type --- stars: kinematics and dynamics --- (stars:) planetary systems --- stars: solar-type --- Galaxy: center --- Galaxy: disk}

\section{Introduction} \label{sec:intro}

It is widely accepted that the Galactic center (GC) of the Milky Way nests a super-massive black hole 
(SMBH) with a mass $M_{\rm SMBH} \sim 4 \times 10^6~ M_{\odot}$ centered around the Sagittarius 
$\rm A^{\star}$ (\sgra) \citep{Ghez1998, ghez2003b, ghez2008, Genzel1997,
Genzel2010, gillessen2009, schodel2009, boehle2016}.  Within 
1 pc$^3$ from the \sgra, there are $\sim 10^7$ mostly mature stars \citep{Do2009, Genzel2010} as well as 
a rich and conspicuous population  ($\sim 80$) of very young massive and Wolf-Rayet stars with 
estimated ages of $4-6$~Myr inside the 
central $\lesssim 0.5$ pc \citep{Ghez2003a}.  At least half of these stars are orbiting around the central SMBH in 
a relatively flat clockwise disk.  These stars are dubbed as clockwise stars (CWSs) \citep{lockmann2009}. 
The possibility of other stars in an inclined  counter-clockwise disk has also been suggested \citep{paumard2006} 
albeit the existence of the second disk remains controversial \citep{lu2006, lu2009}.

It is possible for some mature GC stars to have originated from some stellar clusters which underwent orbital
decay from a few kpc away to their present location within the Hubble time \citep{Tremaine1976a, Gerhard2001, 
Just2011}.  But, the brief (a few Myr) lifespan of the massive main-sequence stars, including the CWSs,  
casts a strong limit on the distance over which they may have migrated \citep{PortegiesZwart2002, 
Gurkan2005, Madigan2014}. It has been suggested that these stars may be formed in a gaseous 
disk around the SMBH \citep{Goodman2003, Levin2003}.  They may 
also have been rejuvenated from old low-mass stars which are trapped and fed by a disk \citep{artymowicz1993}.  Although 
there is no evidence of this disk today, its past presence may be inferred from its subsequent depletion, which 
may be attributed to the outflow from this region in the form of the Fermi bubble \citep{Su2010}.  
These scenarios are compatible with some CWSs' orbits
but not necessarily with all the young stars.  The main outstanding issue for the CWSs is that several of them also have large 
eccentricities ($\sim 0.7$) \citep{lu2006, paumard2006,  gillessen2009, Do2013}.  If these stars acquired 
their masses on low or modest eccentricity ($\lesssim 0.5$) orbits in their natal disk \citep{Nayakshin2007}, 
the time (a few Myr) lapsed since their emergence would not be adequate for local dynamical relaxation 
processes to excite such large eccentricities \citep{Freitag2006, Alexander2007,  Madigan2011}.  It is possible
that some, but not all, of these stars may have originated from a tidally disrupted young cluster 
\citep{Gerhard2001, Kim2003,  Genzel2003, maillard2004, Levin2005}. It is also possible that the CWSs' orbits 
may be perturbed by a compact companion \citep{Hansen2003}, a stellar cluster \citep{Schodel2005} or a second disk
\citep{paumard2006, lockmann2009}.

There are other attempts to explain the orbital properties of these CWSs. In the work of \cite{Naoz2018}, they investigated the critical role of stellar binaries in the GC and found that the existence of binaries could significantly vary the inferred stars’ orbital properties. According to their discussion, the eccentricities of these disk stars in the GC may be less than those
inferred from the kinematic measurement at first glance. In addition, an eccentric nuclear disk could drive new secular instability on stellar orbits in the GC, and amplify deviations of individual orbital eccentricities from the average value \citep{Madigan2009,
Madigan2018}.
Most recently, \cite{Generozov2020} proposed a binary disruption scenario that the S-stars at a  distance of 0.01 pc to the GC are the remnants of tidally disrupted binaries from the CWS population.  Based on this scenario, they can reproduce the observed S-star semi-major axis and thermal eccentricity distribution. 

In this paper, we mainly adopt the perturbation hypothesis with the consideration of an additional effect
which may account the rapid excitation of high eccentricity for stars recently formed in a disk with nearly circular orbits.
There are several observations that impose constraints on the mass and separation of any companion to \sgra. For example, the proper motion measurement of \sgra~against background extragalactic source places an upper mass limit of $\sim 10^4~M_\odot$ on any companion within $10^3-10^5$~AU around the SMBH \citep{Reid2004}. The coincidences of the density as well as the kinematic centers of the young stars with the position of the \sgra~ have also been used to limit the companion's mass to be $\lesssim 10^{-2}~M_{\rm SMBH}$ and separation in the range of $0.01-1$~pc \citep{Yu2003}.

We speculate and assume the culprit to be an infrared-radiation source IRS 13E complex \citep{Krabbe1995} 
which contains some massive and early-type stars. Most of these stars appear to be physically 
bound to the IRS 13E complex \citep{maillard2004} which is located at a projected separation of 
$\sim 0.13$~pc from the \sgra.  Recent ALMA observations also unveiled ionized gas which 
appears flowing along eccentric Keplerian streams around a point-mass potential at 
the location of IRS 13E \citep{tsuboi2017}. However, this interpretation would break down
if the gas properties of IRS 13E is due to a colliding winds \citep{wang2020, zhu2020}.
Although the suggestion that it may be an intermediate-mass black hole (IMBH) remains a 
controversial issue \citep{maillard2004, Hansen2003, Kim2004}, its total inferred mass 
($\sim 10^4 M_\odot$) would be reasonable if the young stars in the IRS 13E complex are 
gravitationally bound \citep{Schodel2005, tsuboi2017}. This estimated total mass is a few
times larger than that from all the massive stars in the region.

The estimated total mass of IRS 13E and its separation from \sgra~ are within the allowable 
parameter space inferred from the observed properties \citep{Yu2003, Reid2004, naoz2020}.  
Moreover, the proper motion of IRS 13E, relative to \sgra, seems to be 
in the same direction as that of stars with counter-clockwise orbits, but opposite to that
of CWS's \citep{Genzel2003}, though large uncertainties remain due to the lack of radial 
velocities and relative distances between it and \sgra.  

Under these circumstances, the dynamical properties of young stars near \sgra, especially in the 
central parsec, are likely to be significantly affected by the secular perturbation from 
such a massive companion.  Here, we suggest the influence of this intermediate-mass companion
(IMC) is particularly strong 
near its secular resonance where its precession frequency matches that of the stars around
\sgra~induced by it.  At the location of its secular resonance, angular momentum is extracted 
from the stars by the IMC such that their eccentricities are largely excited.  During the 
depletion of the CWSs' natal disk, the location of secular resonance sweeps towards \sgra~over 
an extended region, inducing high eccentricity for a large fraction of CWSs. This effect is 
analogous to that induced by Jupiter on the asteroids in the solar system \citep[e.g.,][]{zheng2017a, 
zheng2017b}.  Since secular perturbation is imposed over a large distance and accumulates over
many orbital periods, this IMC does not need to be an IMBH.  A compact and bound stellar cluster 
with a mass $\sim 10^4 M_\odot$ is adequate to provide sufficient secular perturbation.

In this paper, we consider various contributions to the gravitational potential including that
from the super-massive black hole (SMBH), intermediate-mass companion (IMC), the Galactic bulge, disk, and halo
as well as that from a gaseous disk out of which the disk stars emerge (\S\ref{sec:potential}).
In order to analyze the effects of secular interaction, we recapitulate a set of equations analogous
to the Laplace planetary equations in celestial mechanics (\S\ref{sec:planetaryeq}). We utilize these 
equations to analyze the precession due to various components of the gravitational field and the
potential for eccentricity excitation by IRS 13E with various possible orbital configurations. 
Our analytic calculation is verified by a series of numerical-simulation models of secular 
resonances in \S\ref{sec:model}.  Finally in
\S\ref{sec:summary}, we summarize our results and discuss their 
implications.
 
\section{Contributing potentials}
\label{sec:potential}
Our objective is to investigate the statistically orbital distribution (mainly the semi-major axis $a$ and 
eccentricity $e$ of their orbits around \sgra) of CWSs under the influence of the sweeping secular resonance 
of an IMC (i.e., IRC 13E, which may be an IMBH or a very compact stellar cluster). 

\subsection{Numerical method}
To solve the Newtonian equation of motion, we adopt the open-source \textit{N}-body code REBOUND \citep{reinliu2012}, and choose 
the built-in MERCURIUS integrator \citep{rein2019}. In this code, a merger can occur if the radius of two objects overlaps. 
There are several contributors to the potential \citep{Gnedin2005, Widrow2005, Kenyon2008}: the central SMBH ($\Phi_{\rm SMBH}$), 
a satellite IMC ($\Phi_{\rm IMC}$), the Galactic bulge ($\Phi_{\rm bulge})$, 
the contribution from the potential of the Galactic disk and the halo can be negligible in the central parsec.
Within 1 pc from \sgra, the SMBH and IMC have Schwarzchild radii (see Table~\ref{tab:scale}), with $s = 2 G M / c^2$, 
where $c$ is the speed of light. For the SMBH's potential, we include the post Newtonian corrections.
In order to consider the possibility of the IMC being a dense cluster instead of a point mass IMBH,
we also adopt, for one model, a Plummer softening parameter $s^{\prime}_{\rm IMC}$ for the satellite perturber.  
For different components of 
the Galactic potential, we adopt the conventional prescription:
\begin{equation}
\begin{split}
\Phi_{\rm SMBH} & = - \frac{G M_{\rm SMBH}}{r} + {\rm post \ Newtonian \ terms}, \\
 \Phi_{\rm IMC} & = - \frac{G M_{\rm IMC}}{\sqrt{\mid \vec{r}-\vec{r}_{\rm IMC} \mid^2 + s^{\prime 2}_{\rm IMC}}},\\ 
 \Phi_{\rm bulge} & = - \frac{G M_{\rm bulge}}{r+a_{\rm bulge}}, 
 \label {eq:galactic_pot}
\end{split}
\end{equation}
where $r$ is the spherical radius,
$r_{\rm IMC}$ is the distance 
of the IMC from the GC, $a_{\rm bulge}$ is the scaling lengths for the bulge, 
$M_{\rm SMBH}$ and $M_{\rm IMC}$ are the mass of the SMBH and IMC, $M_{\rm bulge}$ is the mass scaling factor for the bulge \citep{Hernquist1990}.
We adopt related parameters from \cite{Gnedin2005}. We assume spherical symmetry for the Galactic bulge 
.  Although tri-axial mass distribution in these components can lead to precessions with long term implications on the secular evolution of single and
binary stars' orbits \citep{Petrovich2017,  Mathew2020}, these effects may be negligible in comparison with the IMC's secular
perturbation.   We introduce a set of mass and length scaling by factors of $4 \times 10^6$ and $10^3$
respectively (see Table~\ref{tab:scale}) to highlight the similarity between the Galactic 
center and the Solar system.  

\begin{deluxetable}{cccc}
\tablenum{1}
\tablecaption{Scaling Galaxy to planetary system\label{tab:scale}}
\tablewidth{0.5pt}
\tablehead{
\nocolhead{} & \colhead{Observation} & \colhead{Scaling factors} & \colhead{Simulation}
}
\startdata
$M_{\rm bulge}$ & $1 \times 10^{10} M_{\odot}$ & $4 \times 10^6$ & $2.5 \times 10^3~M_{\odot}$ \\
$a_{\rm bulge}$ & 0.6 kpc & $1 \times 10^3$ & $1.2 \times 10^5$~AU \\
\\
$M_{\rm SMBH}$ & $4 \times 10^6 M_{\odot}$ & $4 \times 10^6$  & 1 $M_{\odot}$ \\
$s_{\rm SMBH}$ & $8 \times 10^{-2}$~AU & $1 \times 10^3$  &  $8 \times 10^{-5}$~AU \\
\\
$\tau_{\rm dep}$ & $1.58-6.32$~Myr & 15.8  & $0.1-0.4$~Myr \\
$\Sigma_0$ & $600-800 \rm~g/cm^2$ & 4  & $150-200 \rm~g/cm^2$ \\
$R_0$ & 1000 AU & $1 \times 10^3$ & 1~AU \\
$H/R$ & 0.01 & 1 & 0.01 \\
\\
$M_{\rm IMC}$ & $1 \times 10^4 M_{\odot}$ & $4 \times 10^6$ & $2.5 M_{\rm J}$ \\
$s_{\rm IMC}$ & $3 \times 10^{4}$~km & $1 \times 10^3$  &  $30$~km \\
$s_{\rm IMC}^\prime$ & 5000 AU & $1 \times 10^3$  &  $5$~AU \\
$a_{\rm IMC}$ & $0.3-0.4$~pc & $1 \times 10^3$ & $60-80$~AU \\
$e_{\rm IMC}$ & $0.2-0.4$ & 1 & $0.2-0.4$  \\
\\
$m_{\star}$ & $12~M_{\odot}$ & $4 \times 10^6$ & 1 $M_{\oplus}$ \\
$s_{\star}$ & $1.5 \times 10^{6}$~km & $1 \times 10^3$  &  $1.5 \times 10^{3}$~km \\
$a_{\star}$ & $0.05-0.5$~pc & $1 \times 10^3$ & $10-100$~AU \\
$v_{\star}$ & $\sim 200-600~\rm km/s$  &  63 & $\sim 0.1-0.3~v_{\oplus}$ \\
\enddata
\tablecomments{$M_{\rm SMBH}, M_{\rm bulge}$
are the masses of SMBH, Galactic bulge; $a_{\rm bulge}$ is the characteristic length scales
of the Galactic bulge; $M_{\rm IMC}, a_{\rm IMC}, e_{\rm IMC}$ are the mass, semi-major axis and eccentricity of the intermediate mass companion (IMC, which is either an IMBH or stellar cluster); $s_{\rm SMBH}, s_{\rm IMC}$ are 
the Schwarzchild radii of the SMBH and IMC; $s_{\rm IMC}^\prime$ is
the softening parameter to represent IRS 13E as a stellar cluster, $\Sigma_0, H/R, \tau_{\rm dep}$ 
are the reference surface density, aspect ratio, characteristic depletion time of the gaseous 
disk, $m_\star, s_\star, a_\star, v_\star$ are the mass, radius, initial semi-major axis, and 
velocity of the stars.}
\tablecaption{}
\end{deluxetable}
In this paper, we focus our attention on the excitation of disk stars within 0.5 pc.  Gravity due to
the Galactic disk and halo has negligible influence on the dynamical evolution of these stars and the IMC.
In paper II of this series, we show that the Galactic disk and halo potentials can significantly modify 
the trajectories of the unbound hyper-velocity stars at large distances ($>$ kpc) from the Galactic center.

The gravitational potential of a hypothetical geometrically thin gaseous disk (in which the young stars emerge) 
is determined by its poorly constrained surface density ($\Sigma$) distribution.  We adopt a prescription similar 
to our previous models of protostellar disks \citep{nagasawa2005, zheng2017a}.  
For illustrative purpose, we assume a power-law surface density ($\Sigma$) distribution constructed for the minimum mass solar nebula
\citep{Hayashi1985} such that
\begin{equation}
    \Sigma = \Sigma_0 \left(\frac{r}{R_0} \right)^{-3/2} e^{-t/\tau_{\rm dep}} ~\rm g/cm^2,
\end{equation}
where $\tau_{\rm dep}$ is the depletion timescale of the gas nebula, $R_0$ is a fiducial cylindrical radius. We take into account the possibility 
that the IMC can strongly perturb the structure of the disk near its orbit analogous to proto giant planets in 
protostellar disks.  But the much less massive young stars have negligible effects on the disk structure, 
analogous to planetesimals in a protostellar disk.  For models in which the IMC's orbital angular momentum vector
is parallel to that of the disk (co-orbiting models), we assume the presence of a gap around its orbit at
\begin{equation}
a_{\rm in/out} = a_{\rm IMC} (1 \mp e_{\rm IMC}) \left[ 1 \mp \left( \frac{M_{\rm IMC}}{3M_{\rm SMBH}}\right)^{1/3} \right], 
\label{eq:ainout}
\end{equation}
where $a_{\rm IMC}$ and $e_{\rm IMC}$ are the orbital parameters of the IMC.  The structure of this gap can modify the
disk's force on the IMC and precession rate of its orbit \citep{ward1981, nagasawa2005, zheng2017a} such that
\begin{equation}
\begin{split}
& f_{\rm gap, gas}   =  2\pi G \Sigma \sum_{l=0}^{\infty} \left[ \frac{(2l)!}{2^{2l} (l!)^2} \right]^2 
\left(\frac{2}{4l+1}\right)  \\
& \times \left[ (2l) \left( \frac{r}{a_{\rm out}}  \right)^{2l+1/2}  - (2l+1) \left( \frac{a_{\rm in}}{r} \right)^{2l+1/2}  \right] .
\end{split} 
\label{eq:f_gas_IMC}
\end{equation}
For the case where we consider the possibility that the IMC's orbital angular momentum vector is anti-parallel to that of the disk
(counter-orbiting models), the IMC does not open up a gap in the gaseous disk.  We also assume the embedded stars do not have 
sufficient mass to open gaps near their orbit and their apsidal precession induced by the gas disk is dominated by the gas in 
their neighborhood.  The disk's gravitational force on them can be computed as 
\begin{equation}
f_{\rm emb, gas} = - 4\pi G \Sigma Z_{k} ,
\label{eq:f_gas_star}
\end{equation}
where $Z_k = 1.094$ is the same as in \cite{ward1981, nagasawa2005, zheng2017a}.

In addition, a post-Newtonian correction for the central SMBH is included, using the relativity formula from \cite{kidder1995,nagasawa2005}, in the form as
\begin{equation}
\begin{split}
 \vec{f}_{\rm rela}  &= \frac{G M_{\rm tot}}{r^2 c^2} 
 \Bigg\{ 2(2-\mu)v_{r} \vec{v} -  \bigg[ (1+3\mu)v^2  \\
  &   - 2(2+\mu)\frac{G M_{\rm tot}}{r} -\frac{3}{2}\mu v_{r}^{2} \bigg] \frac{\vec{r}}{r} \Bigg\}  ,
\end{split}
\end{equation}
where $v_r$ is the velocity in the $r$ direction, and $M_{\rm tot} = M_{\rm SMBH}+m$, $\mu = m M_{\rm SMBH}/M_{\rm tot}^2$. 
Note that it is just a simple $1$-post-Newtonian form. Motivated by the work of \cite{Rodriguez2018}, we have made a further test on the $2$-post-Newtonian term. We find that the higher-order correction makes negligible correction on most residual orbits, except for those highly eccentric orbits, which have the closest distance to the central SMBH smaller than $\sim 10^{-5}$~pc. Therefore, our assumption concerning general relativity is appropriate in this work. In the second paper of this series, we extend this formula to include the higher-order term \citep{kidder1995} to avoid considerable energy alterations during close encounters with the SMBH.

\begin{deluxetable*}{cccccc}
\tablenum{2}
\tablecaption{Parameters for various numerical simulations \label{tab:models}}
\tablewidth{0pt}
\tablehead{
\colhead{models} 
&\colhead{${s^{\prime}_{\rm IMC} (\rm AU)}^a$} &\colhead{$i_{\rm IMC, \star} (^{\circ})^b$}  &\colhead{$i_{\star}(^{\circ})^c$}
&\colhead{${\rm Gas \ Disk \ Potential}^{d}$}  &\colhead{${\rm Gas \ Disk \ Gap}^e$}  }   
\startdata
NGD         &  0 & 0  & 0   &  $-$ &  $-$ \\
fiducial    &  0 & 0 & 0   &  $+$  & $+$ \\
SOFT         &  5000 & 0  & 0   &  $+$ & $+$ \\
CCW         &  0 & 180 &  0   & $+$ &  $-$ \\
CCW-NGD      &  0 & 180 & 0  &  $-$ &  $-$ \\
I30           &  0 &  30  & 0  &  $+$  & $-$ \\
I30-NGD      &  0 &  30  & 0  &  $-$  & $-$ \\
I60          &  0 &  60  & 0  &  $+$  & $-$ \\
I60-NGD      &  0 &  60  & 0  &  $-$  & $-$ \\
I60-IS60     &  0 &  60  & 60  &  $+$  & $-$ \\
I60-IS60-NGD   &  0 &  60  & 60  &  $-$  & $-$ \\
\enddata
\tablecomments{ $^a$ The softening parameter of IMC's potential. $^b$ The initial mutual inclination between the orbits of IMC and disk stars. $^c$ The initial inclination of the orbit of disk stars (gas disk) relative to the Galactic disk (reference plane).$^d$ The presence of the gas disk potential.$^e$ The presence of a gas-free gap around the IMC's orbit. }
\end{deluxetable*}

\subsection{Numerical models}
In order
to investigate the effects of sweeping secular resonance on the evolution of GC stars and its parameter dependence, we test a series of 
numerical models with various setups. Table \ref{tab:models} lists all these models, including simulations with three different gas-disk
depletion timescales.  In all cases, the stars are assumed to reside in the gaseous disk with circular orbits initially.  
In the proximity of the SMBH, the potential contribution from the Galactic disk is negligible.  For simplicity, we consider the 
case that the disk stars' initial orbits are in the same plane as that of the Galactic disk.  Contributions from the Galactic disk and
halo are important for the hypervelocity stars after they are ejected from the proximity of the SMBH.  An inclination angle between the 
disk stars' orbits and the Galactic disk may introduce some modification to the asymptotic kinematic properties of the HVSs, we will discuss this circumstance in paper II of this series.


We highlight the influence of the sweeping secular resonance by turning off gravity due to the gas in a comparative model NGD. 
In all but one model, the companion's potential is assumed to be that of a point mass IMBH. Since the IMBH nature of IRS 13E continues 
to be a controversial issue, we also consider the possibility of the IMC being a dense star cluster with a Plummer-type mass 
distribution in the SOFT model with a non-zero $s_{\rm IMC}^{\prime}$ (see Table~\ref{tab:models}). 

In the fiducial, NGD, and SOFT models, we assume the \sgra, companion, stars and gaseous disk are all in the same plane,
and their orbital angular momentum vectors are aligned. In the CCW model, we consider the possibility that the IMC's orbit 
is anti-parallel to that of the stars and gaseous disks, as suggested by some observations of IRS 13E \citep{Genzel2003}.
In this paper, we present the analysis and simulations of these co-planar models.
We also consider a series of 3-d cases with various misalignment angles between
the IMC's and the stars' initial orbital planes.  For those simulations, we approximate the
gaseous disk's force in the radial direction of the gaseous disk with 
$f_{\rm{emb, gas}} (R_{\rm IMC})$ (Equation~(\ref{eq:f_gas_star})) in models I30 and I60 to take into account the 
possible absence of a gap at the 
gaseous disk radial location $R_{\rm IMC}$.  For both models, the force in the ($z$, the distance above the disk) direction 
normal to the gaseous disk plane is approximated with 
\begin{equation}
    f_{\rm z, gas}  = - 2\pi G \Sigma(R) \frac{ {\rm sign}(z) (z/H)^2}{1 + (z/H)^2} .
\end{equation}
The gaseous disk's aspect ratio $H/R$ is assumed to be 0.01 (see Table~\ref{tab:scale}) at $R_0$, with the form 
$H/R = (H_0/R_0) (R/R_0)^{1/4}$.  The gravitational stability parameter of the disk is
\begin{equation}
    Q = \frac{c_s \Omega}{\pi G \Sigma} = \frac{H_0}{R_0} \frac{M_{\rm SMBH}}{\pi \Sigma_0 R_0^2} \left( \frac{R}{R_0} \right)^{-1/4} \sim {\rm a~few~} 10  ,
\end{equation}
such that the effects of self gravity is negligible in these models.  
For comparison, we also examine the cases without the gaseous disk.

\section{Celestial mechanics of a representative star}
\label{sec:planetaryeq}
In the numerical simulations, we integrate the stars' orbits based on the Newtonian equation of motion.
In the analysis of this process, it is useful to adopt Laplace's approach in celestial mechanics by
averaging periodic velocity and position changes over an orbital timescale. We 
consider the secular (long-term) evolution of orbital elements, including their semi-major axis $a$, 
eccentricity $e$, inclination $i$, longitudes of the periastron $\varpi$, and ascending node $\Omega$.
This approach provides physical insights into the dominant contribution of various effects.  

\subsection{Secular interaction between IMC and the disk stars}
For stars reside within the inner parsec from the \sgra, the predominant contributor to the gravitational is 
$\Phi_{\rm SMBH}$ and other components can be treated as perturbations. This configuration is closely analogous
to solar system dynamics, where the secular evolution in celestial mechanics can be approximated by
planetary equations \citep{murray2000}.  We first construct a set of disturbing function
\begin{equation}
\begin{split}
    \Re  &= \Re^{(\rm sec)} + \Re^{(\rm bulge)} + \Re^{(\rm gas)} + \Re^{(\rm GR)}.
 \end{split}
\end{equation}
%
In addition, general relativity (GR) is only significant 
in the limit of small periastron distance, $r_p = a(1-e)$, see also in Figure~\ref{fig:a_t}. For most of disk stars and the IMC at distance $\sim 0.05-0.5$~pc,  
we neglect the GR precession in our analytic approximation.
 
The secular perturbation between the disk star and the IMC can be expressed as
\begin{equation}
\begin{split}
    \Re_{\star}^{(\rm sec)} &= G M_{\rm IMC} \left( \frac{1}{\mid \vec{r}_{\rm IMC} - \vec{r}_{\star} \mid} - \frac{\vec{r}_{\star} \cdot \vec{r}_{\rm IMC}}{r_{\rm IMC}^3} \right)  , \\
    \Re_{\rm IMC}^{(\rm sec)} &= G m_{\star} \left( \frac{1}{\mid \vec{r}_{\star} - \vec{r}_{\rm IMC} \mid} - \frac{\vec{r}_{\star} \cdot \vec{r}_{\rm IMC}}{r_{\star}^3} \right)  .
    \end{split}
    \end{equation}
Since $m_{\star} << M_{\rm IMC}$, the secular perturbation of disk stars on the IMC, $\Re_{\rm IMC}^{(\rm sec)}$, can be ignored. 

When we average over an orbit, only the perturbation function's secular term does not contain mean longitude remain.  We adopt the second-order expansion of the direct part of the disturbing function following Equation ($7.8-7.12$) in the \cite{murray2000}, as
\begin{equation}
\begin{split}
  \Re_{\star}^{(\rm sec)}  &=  n_{\star} a_{\star}^2 \bigg[\frac{1}{2} A_1 e_{\star}^2 + A_2 e_{\star} e_{\rm IMC}  {\rm cos}(\varpi_{\star} - \varpi_{\rm IMC})   \\
    & -  2 A_1 {\rm sin}^2\left(\frac{i_{\star}}{2} \right) +  4 A_1 {\rm sin}\left(\frac{i_{\star}}{2} \right) {\rm sin} \left(\frac{i_{\rm IMC}}{2}\right) \\
    & \times {\rm cos}(\Omega_{\star} - \Omega_{\rm IMC}) \bigg] , \\
 {\rm with} \ \ & A_1 = \frac{n_{\star} \mu_{\star} \alpha^2 }{4} b_{3/2}^{(1)}  \ \ {\rm and}
\  A_2 = - A_1 \frac{b_{3/2}^{(2)}}{b_{3/2}^{(1)}} ,
 \end{split}
 \end{equation}
where ($a_{\star}$, $e_{\star}$, $i_{\star}$, $\varpi_{\star}$, $\Omega_{\star}$, $n_{\star}$) and ($a_{\rm IMC}$, $e_{\rm IMC}$, $i_{\rm IMC}$, $\varpi_{\rm IMC}$, $\Omega_{\rm IMC}$, $n_{\rm IMC}$) are the semi-major axis, eccentricity, inclination, longitudes of the peri-astron, ascending node, mean angular velocity of the disk star and the IMC, respectively. We have assumed that $a_{\star} < a_{\rm IMC}$ as most stars are located within the orbit of the IMC.  In the limit $m_{\star} < < M_{\rm IMC} < <M_{\rm SMBH}$,  $n_{\star} \approx \sqrt{G M_{\rm SMBH} / a_{\star}^3}$ , $\mu_{\star} \approx M_{\rm IMC}/M_{\rm SMBH}$, 
$b_{3/2}^{(1)}$ and $b_{3/2}^{(2)}$ are Laplace coefficients for the semi-major axis ratio $\alpha \equiv a_\star/a_{\rm IMC}$.

The stars' Lagrangian equations of motion are
\begin{equation}
\begin{split}
   \frac{d \varpi_{\star}}{d t}^{\rm (sec)}  &= \bigg[A_1 + A_2 \left( \frac{e_{\rm IMC}}{e_{\star}} \right) {\rm cos}(\varpi_{\star} - \varpi_{\rm IMC}) \bigg]\sqrt{1-e_{\star}^2}  \\
  & - 2 A_1 \bigg[ {\rm sin}^2 \left(\frac{i_\star}{2} \right) -  {\rm sin} \left(\frac{i_\star}{2} \right) {\rm sin} \left(\frac{i_{\rm IMC}}{2} \right)  \\
  & \times {\rm cos}(\Omega_{\star} - \Omega_{\rm IMC})  \bigg]  \frac{1}{\sqrt{1-e_{\star}^2}}   , \\
 \frac{d e_{\star}}{d t}^{\rm (sec)} &=  A_2 e_{\rm IMC} {\rm sin}(\varpi_{\star} - \varpi_{\rm IMC})\sqrt{1-e_{\star}^2}   .
\end{split}
\end{equation}
In the limit of small $e_\star$, we approximate $\sqrt {1 - e^2_\star} \simeq 1$.  Since the disk stars' secular perturbation 
on the IMC is negligible 
\begin{equation}
\frac{d e_{\rm IMC}}{d t}^{\rm (sec)} \simeq \frac{d \varpi_{\rm IMC}}{d t}^{\rm (sec)} 
\simeq 0.
\end{equation}

\subsection{Precession due to the bulge and other components of the Galactic potential}

The bulge also induces disk stars and IMC to precess at rates comparable to that of the
stars under the IMC's secular perturbation.  At locations $r  << a_{\rm bulge}$,  the 
disturbing function due to perturbation from the potential of the Galactic bulge can be written as
\begin{equation}
\begin{split}
\Re^{(\rm bulge)} = \frac{G M_{\rm bulge}}{a_{\rm bulge}} \left(1 - \frac{r}{a_{\rm bulge}} \right) .
\end{split}
\end{equation}
This disturbing function can be further simplified without the $r$ independent term.

Although the bulge potential is assumed to be spherically symmetric,
the stars' eccentricity (relative to the SMBH and is defined by the Kepler's equation) varies along the orbit because 
the bulge's potential is contributed by a distributed (rather than a point) mass.  Nevertheless, after averaging 
over the stars' orbits, the Lagrange's equations for the orbital elements are reduced to 
\begin{equation}
\frac{d \varpi} {d t}^{\rm (bulge)} = -\frac{G M_{\rm bulge}}{a_{\rm bulge}^2}  \frac{\sqrt{1-e^2}}{n a}
\end{equation}
and $(d e/d t)^{\rm (bulge)} =0$ because the potential of a spherically symmetric bulge does not
induce any net changes in the star's orbital energy and angular momentum on a secular timescale.

Note that the precession rate is relative to the orbits of either the disk stars or the IMC.  Since 
$(d \varpi/d t)^{\rm (bulge)} <0$, the precession
of stars/IMC on a clockwise/counter-clockwise orbit is counter-clock/clockwise. Within 1 pc$^3$ from the \sgra, we neglect 
contributions to the potential from the Galactic disk and halo because they are much weaker than those due to 
IMC and the Galactic bulge.

\subsection{Precession due to the gaseous-disk's potential}
The precession rate of the periastron's longitude induced by the gas disk can be classified as two types according to various 
situations. Disk stars generally do not have sufficient mass to open a gap in the proximity of their orbit.  Unless they are 
very close to the IMC, their precession rate can be derived from Equation~(\ref{eq:f_gas_star}) such that
\begin{equation}
\frac{d \varpi} {d t}^{\rm (emb, gas)} \approx  - \frac{\pi G \Sigma(a) Z_k}{n a} .
\end{equation}
If the IMC orbits around the SMBH in the opposite direction as the gaseous disk, it would not be able to open a gap such that 
is precession rate would have the same magnitude but with an opposite sign. 
 
In contrast, if the IMC revolves around the SMBH in the same direction as the gas disk and the disk stars, it would be able to
open a gap (Equation~(\ref{eq:ainout})).  Under such a circumstance, we can derive its precession rate from Equation~(\ref{eq:f_gas_IMC}), 
such that
\begin{equation}
\begin{split}
& \frac{d \varpi_{\rm IMC}} {d t}^{\rm (gap, gas)}  \approx  \frac{ \pi G \Sigma(a_{\rm IMC})}{n_{\rm IMC} a_{\rm IMC}} Z_{\rm IMC}  \ \ \ \ \ \ {\rm where}  \\
&  Z_{\rm IMC} =  \sum_{l=1}^{\infty} \left[ \frac{(2l)!}{2^{2l} (l!)^2} \right]^2 \frac{4l(2l+1)}{4l+1} \\
& \times \left[  \left( \frac{a_{\rm IMC}}{a_{\rm out}}  \right)^{2l+1/2} 
+  \left( \frac{a_{\rm in}}{a_{\rm IMC}} \right)^{2l+1/2}  \right] . 
\end{split}
\end{equation}
Due to the presence of the gap, IMC's direction of precession is opposite to that of the disk stars \citep{nagasawa2005, ward1981}. 
Similar to $(d \varpi/d t)^{\rm (bulge)}$, negative value of  $(d \varpi/d t)^{\rm (emb, gas)}$ implies the 
periastron longitude' regression relative to the sense of stars'/IMC's orbit.
Here we neglect the effect of eccentricity damping by either tidal or hydrodynamic drag and 
assume $(da/dt)^{\rm (gas)}=(de/dt)^{\rm (gas)} =0$.

\subsection{Relative longitude of periasteron between the disk stars' and co-orbiting or 
counter-orbiting IMC} 
The orbital evolution of both the disk stars and IMC can be computed by the sum of all contributions.
For simplicity, we only analyze the configuration that the disk stars, the gaseous disk, and the IMC are 
coplanar.  (The non-coplanar models are simulated numerically in the next section.)
There are two limiting co-orbiting and counter-orbiting cases, in which the orbital angular 
momentum vector of the IMC is either
a) parallel or b) anti-parallel with respect to that stars and their natal gaseous disk.
Without the loss of generality, we can set $i_\star=0$ so that $i_{\rm IMC}=0$ corresponds to
the co-orbiting case and $i_{\rm IMC}=\pi$ to the counter-orbiting case.  
In both cases, the precession rate of the disk stars is

\begin{equation}
\begin{split}
\frac{d \varpi_{\star}}{d t} &=  \frac{d \varpi_{\star}}{d t}^{(\rm sec)} + \frac{d \varpi_{\star}}{d t}^{(\rm bulge)}  + \frac{d \varpi_{\star}}{d t}^{(\rm emb, gas)} \\
& \simeq  A_1 + A_2 \left(\frac{e_{\rm IMC}}{e_\star} \right) {\rm cos} (\varpi_\star - \varpi_{\rm IMC}) \\
& -  \frac{G M_{\rm bulge}/a_{\rm bulge}^2}{ n_{\star} a_{\star}}     - \frac{\pi G \Sigma(a_{\star}) Z_k}{n_{\star} a_{\star}}  . 
\end{split}
\label{eq:dvarpistar}
\end{equation}

The co-orbiting IMC  (with $i_{\rm IMC}=i_\star =0$) can open a gap in the gaseous disk such that its precession rate becomes
\begin{equation}
\begin{split}
\frac{d \varpi_{\rm IMC}}{d t} &=  \frac{d \varpi_{\rm IMC}}{d t}^{(\rm bulge)}  + \frac{d \varpi_{\rm IMC}}{d t}^{(\rm gap, gas)}  \\ 
&\approx  -  \frac{G M_{\rm bulge} / a_{\rm bulge}^2}{ n_{\rm IMC} a_{\rm IMC}}  \   +\frac{ \pi G \Sigma(a_{\rm IMC})
Z_{\rm IMC}}{n_{\rm IMC} a_{\rm IMC}}    ,
\end{split}
\label{eq:dvarpiIMC1}
\end{equation}
We obtain, in the limit of small $e_\star$, the evolution of $\xi \equiv e_\star/ e_{\rm IMC}$ and
the angle between IMC's and the stars' longitude of periastron $\eta=\varpi_\star - \varpi_{\rm IMC}$ 
from Equations (\ref{eq:dvarpistar}) and (\ref{eq:dvarpiIMC1}),
\begin{equation}
\begin{split}
     \frac{d \eta}{dt} & \simeq  A_{\rm tot} + \frac{A_2}{\xi} {\rm cos}~\eta  \ \ \ \ \ \
  {\rm and} \ \ \ \ \ \  \frac{d \xi}{dt} \simeq A_2 {\rm sin}~\eta, \\
   A_{\rm tot} &=  A_1 - \frac{G M_{\rm bulge}}{a_{\rm bulge}^2} \left( \frac{1}{n_\star a_\star} - \frac{1}{n_{\rm IMC} a_{\rm IMC}} \right) \\
&  - \pi G \left( \frac{\Sigma (a_\star) Z_k}{ n_\star a_\star} + \frac{\Sigma (a_{\rm IMC} ) Z_{\rm IMC}}{n_{\rm IMC} a_{\rm IMC}} \right).
\end{split}
\label{eq:prograde}
\end{equation}
In the above expression, we have taken into account the effect of a gap is to induce IMC to precess in the opposite direction
as the stars.  

Within an order of magnitude,
\begin{equation}
\begin{split}
   &   \frac{\vert {d \varpi_{\star}}/{d t}^{(\rm bulge)} \vert}{\vert {d \varpi_{\star}}/{d t}^{(\rm sec)} \vert}
     \simeq \mathcal O (4 B_1), \\
   &   \frac{\vert {d \varpi_{\star}}/{d t}^{(\rm gas)} \vert}{\vert {d \varpi_{\star}}/{d t}^{(\rm sec)} \vert}
     \simeq \mathcal O \left( \frac{B_2 Z_k}{\alpha^{3/2} {\rm exp} (t/\tau_{\rm dep})} \right), \\
 &    \frac{\vert {d \varpi_{\rm IMC}}/{d t}^{(\rm bulge)} \vert}{\vert {d \varpi_{\star}}/{d t}^{(\rm sec)} \vert}
    \simeq \mathcal O \left( \frac{4 B_1}{\alpha ^{1/2}} \right), \\
 &      \frac{\vert {d \varpi_{\rm IMC}}/{d t}^{(\rm gas)} \vert}{\vert {d \varpi_{\star}}/{d t}^{(\rm sec)} \vert}
    \simeq \mathcal O \left(  \frac{B_2  Z_{\rm IMC}}{ \alpha^{1/2} {\rm exp} (t/\tau_{\rm dep}) } \right), \\
& {\rm with} \ \ \ \ B_1 \equiv 
    \frac{M_{\rm bulge} a_{\rm IMC}^2}{M_{\rm IMC} a_{\rm bulge}^2}, \ \ B_2 \equiv \frac{M_D a_{\rm IMC} ^{1/2}}{ M_{\rm IMC}  R_0^{1/2} },
    \end{split}
    \label{eq:pratio}
\end{equation}
where $M_D = 4 \pi \Sigma_0 R_0^2$ is the characteristic disk mass.
Different components of $A_{\rm tot}$ have similar magnitude at $t=0$ and the contribution from the gaseous
disk decreases with time as it is depleted.  
 
A counter-orbiting IMC (with $i_{\rm IMC}= \pi$ and $i_\star =0$) cannot open up a gap in 
the stars' natal gaseous disk.  In this case, $\varpi_{\rm IMC}$ and
$\varpi_\star$ are measured in the opposite sense.  In the determination of $\eta$ for the IMC's secular
perturbation on the disk stars, we need bring the values of $\varpi_{\rm IMC}$ to the same frame where 
$\varpi_\star$ is measured by switching its sign such that $\eta= \varpi_\star+\varpi_{\rm IMC}$ and  
\begin{equation}
\begin{split}
    \frac{d \eta}{dt} & \simeq A_{\rm tot} + \frac{A_2}{\xi} {\rm cos}~\eta  \ \ \ \ \ \
  {\rm and} \ \ \ \ \ \  \frac{d \xi}{dt} \simeq A_2 {\rm sin}~\eta, \\
  A_{\rm tot} & = A_1 - \frac{G M_{\rm bulge}}{a_{\rm bulge}^2} \left( \frac{1}{n_\star a_\star} + \frac{1}{n_{\rm IMC} a_{\rm IMC}} \right) \\
&  - \pi G \left( \frac{\Sigma (a_\star ) Z_k}{n_\star a_\star} + \frac{\Sigma (a_{\rm IMC} )Z_k}{n_{\rm IMC} a_{\rm IMC}} \right).
\end{split}
\label{eq:retrograde}
\end{equation}

\subsection{Relative importance of the Galactic bulge, gaseous disk, and IMC's orbit} 
Equations (\ref{eq:prograde}) and (\ref{eq:retrograde}) are the governing equations for disk stars' eccentricity
excitation and nodal precession.  The characteristic timescale can be normalized as a dimensionless independent
variable
\begin{equation}
    \tau =  \frac{M_{\rm IMC}}{M_{\rm SMBH}} \frac{n_{\rm IMC} t}{4}
\end{equation}
\begin{equation}
    \frac{d \eta}{d\tau} \simeq {A}_{\rm tot} ^\prime 
    - \frac{\alpha^{1/2} b^{(2)} _{3/2}}{\xi}
    {\rm cos}~\eta, 
   \ \ \ \frac{d \xi}{d \tau} = - \alpha^{1/2} b^{(2)} _{3/2} {\rm sin}~ \eta,
    \label{eq:dimensionless}
\end{equation}
\begin{equation}
\begin{split}
    {A}_{\rm tot} ^\prime &= \alpha^{1/2} b^{(1)} _{3/2} 
    - 4 B_1 ( \alpha^{1/2} \mp 1 )   \\
   & - \frac{B_2}{{\rm exp} (t/\tau_{\rm dep})}
    \left( \frac{Z_k}{\alpha} + Z_{{\rm IMC}, k} \right).
    \label{eq:atotprime}
\end{split}
\end{equation}
The negative/positive signs inside the bracket of the second term on the
right hand side of Equation (\ref{eq:atotprime}) and 
the IMC and k subscript of $Z$ refer to co-orbiting and counter-orbiting
IMC respectively.

The rate and amount of eccentricity excitation of the disk stars (Eq. \ref{eq:dimensionless}) is
determined by the model parameters $\alpha$, $B_1$ and $B_2$,
Based on our adopted model parameters (Table \ref{tab:scale}), with $\Sigma_0 = 600~\rm{g/cm^2}$,
\begin{equation}
\begin{split}
& B_1 = 0.25, \ \ \ \ \ \ B_2 = 0.66 \ \ \ \ \ \ {\rm for} \ a_{\rm IMC} = 0.3~ {\rm pc},\\
& B_1 = 0.44, \ \ \ \ \ \ B_2 = 0.76 \ \ \ \ \ \ {\rm for} \ a_{\rm IMC} = 0.4~ {\rm pc}.
\label{eq:B1B2}
\end{split}
\end{equation}
The Laplace coefficients $b^{(1)} _{3/2} (\alpha=0.3)=1.07$, 
$b^{(1)} _{3/2} (0.5)=2.58$, $b^{(2)} _{3/2} (0.3)=0.4$,
and $b^{(2)} _{3/2} (0.5)=1.56$. And $Z_{\rm IMC}$ = 1.3.
With this range of values, the contribution from IMC's secular perturbation,
Galactic bulge, and gaseous disk have comparable influence.  

\subsection{Excitation of disk stars' eccentricity due to IMC and Galactic bulge}
\label{sec:nogas}

For simplicity, we fist consider the contribution of the IMC and Galatic bulge
and set $\Sigma_0=0$ so that $B_2=0$.  We adopt $a_{\rm IMC} = 0.3$~pc, $e_{\rm IMC}
=0.2$, and 
$a_\star =0.15$~pc (i.e. $\alpha=0.5$) as a typical illustrative example.
The magnitude of 
\begin{equation}
 A^\prime _{\rm tot} \simeq 2.1 \ \ \ \ \ \ {\rm or} \ \ \ \ \ \    A^\prime _{\rm tot} \simeq 0.1
\label{eq:atotbulge}
\end{equation}
for co-orbiting and counter-orbiting IMC respectively.

We assume that the disk stars formed with negligible 
$e_\ast$ and $\xi = e_\ast/e_{\rm IMC} < < 1$ such that $\vert A ^\prime _{\rm tot} \vert 
< < \vert \alpha^{1/2} b^{(2)} _{3/2} /\xi\vert$.  While the IMC precesses at its own pace 
(determined by $d \varpi_{\rm IMC} ^{\rm (bulge)} / dt)$, $\eta$ evolves to a critical 
non-zero value 
\begin{equation}
    \eta_{\rm crit}= {\rm cos}^{-1} \left( \frac{A^\prime _{\rm tot} \xi}{\alpha^{1/2} b^{(2)} _{3/2}} \right) 
    \label{eq:etacrit}
\end{equation}
such that $d \eta/dt \simeq 0$.  For $\xi < < 1$, $\eta_{\rm crit} \simeq -\pi/2 +\epsilon$ where $0<\epsilon < < 1$.

Under this secular resonance (a match between the IMC's and the stellar precession
rates), $\eta_{\rm crit}$ monotonically increases to $0$ and $e_\ast$ (and $\xi$) increases 
at slowly-declining rate towards a critical value
\begin{equation}
    \xi_{{\rm limit}} \simeq \frac{\alpha^{1/2} b^{(2)} _{3/2}}{\vert A_{\rm tot}^{\prime} \vert}
    \ \ \ \ \ \ {\rm and} \ \ \
    \ \ \  e_{\star, {\rm limit}} = e_{\rm IMC} \xi_{{\rm limit}}
    \label{eq:elimit}
\end{equation}
on a characteristic timescale 
\begin{equation}
    \tau_{\rm ecc} \simeq   \frac{4 M_{\rm SMBH}}{n_{\rm IMC} M_{\rm IMC} \vert A^\prime _{\rm tot} \vert}.
    \label{eq:tauecc}
\end{equation}  
Substitute $\xi_{\star, {\rm limit}}$ and $\tau_{\rm ecc}$ into Equation (\ref{eq:dimensionless}), we find
\begin{equation}
    \tau_{\rm ecc} \frac{\partial \eta}{\partial t}  = 1 - \frac{\xi_{{\rm limit}}}{\xi} {\rm cos}~\eta, 
    \ \ \ \ \ \ \tau_{\rm ecc} \frac{\partial}{\partial t} \left( \frac{\xi}{\xi_{{\rm limit}}} \right)= - 
    {\rm sin}~\eta.
\end{equation}
Provided $e_{\star, {\rm limit}} < 1$, there is a transition from monotonic eccentricity ($e\star$ and $\xi$)
growth to a state of liberation (with $\eta$ around 0 and $\xi$ around $\xi_{{\rm limit}}$). 
The liberation period is $\sim \tau_{\rm ecc}/ 2 \pi$ and the liberation amplitude of $\xi$ is limited
to $\sim \xi_{{\rm limit}}$.  Stars with $\xi \gtrsim 2 ~\xi_{{\rm limit}}$ circulate with limited amplitude in 
$\xi$ and $(-\pi,\pi)$ in $\eta$.

If the IMC revolves around the SMBH in the same direction as the disk stars, $e_{\star, {\rm limit}} 
\sim 0.1$ within $\tau_{\rm ecc} \sim 1$~Myr.  But, if
the IMC revolves around the SMBH in the opposite direction as the disk stars, 
$e_{\star, {\rm limit}} (> 1)$ cannot be reached before $e_\star$ approaches unity on a timescale 
\begin{equation}
    \tau_{\rm parabolic} \simeq   \frac{4 (M_{\rm SMBH}/M_{\rm IMC})}{n_{\rm IMC} \alpha^{1/2} b^{(2)}_{3/2} e_{\rm IMC}} \lesssim 10~ 
    {\rm Myr}.
    \label{eq:tparabolic}
\end{equation}  
In this case, the stars' periastron can reach the proximity of the SMBH and be ejected as high 
velocity stars, although we need to take into account the factor of $\sqrt {1-e_\star ^2}$ and relativistic 
precession in quenching
the evolution of both $d \eta/dt$ and $d \xi/dt$ as $e_\star$ approaches unity. 

\begin{figure*}[ht!]
\epsscale{0.6}
\plotone{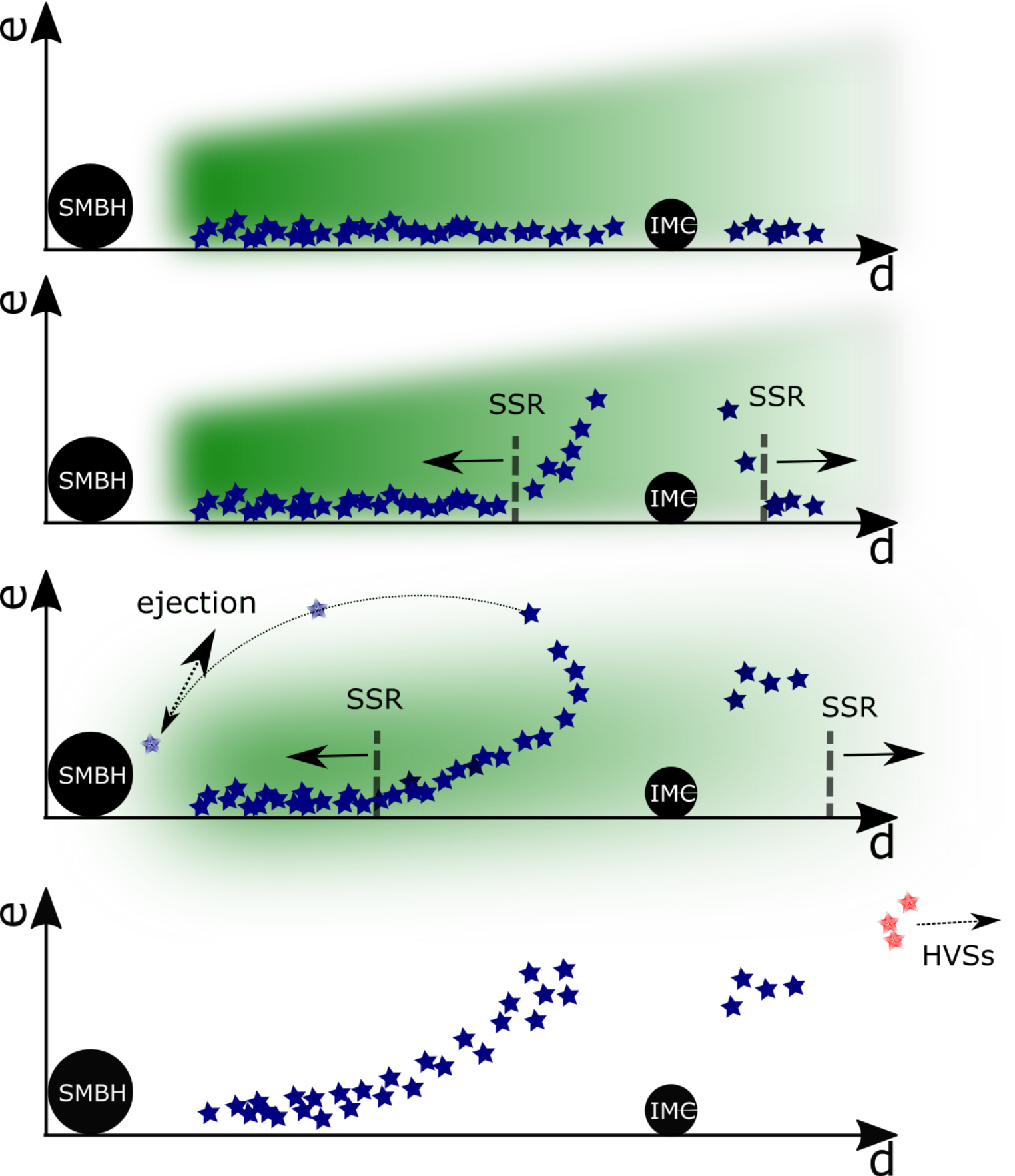}
\caption{A schematic of the effects of the sweeping secular resonance (SSR) on the surrounding disk stars. The GC SMBH and an IMC are labeled with black dots, while blue stars represent the clockwise disk stars. The green shading represents a depleted gas disk, and its color dilutes with the decrease of the gas surface density. In the bottom panel, some hyper-velocity stars (HVSs) are shown with red stars. Their properties are studied in detail in paper II of this series. 
\label{fig:scenario}}
\end{figure*}

\subsection{Nature of sweeping secular resonance}
\label{sec:SSR}

\begin{figure}[ht!]
\epsscale{1.2}
\plotone{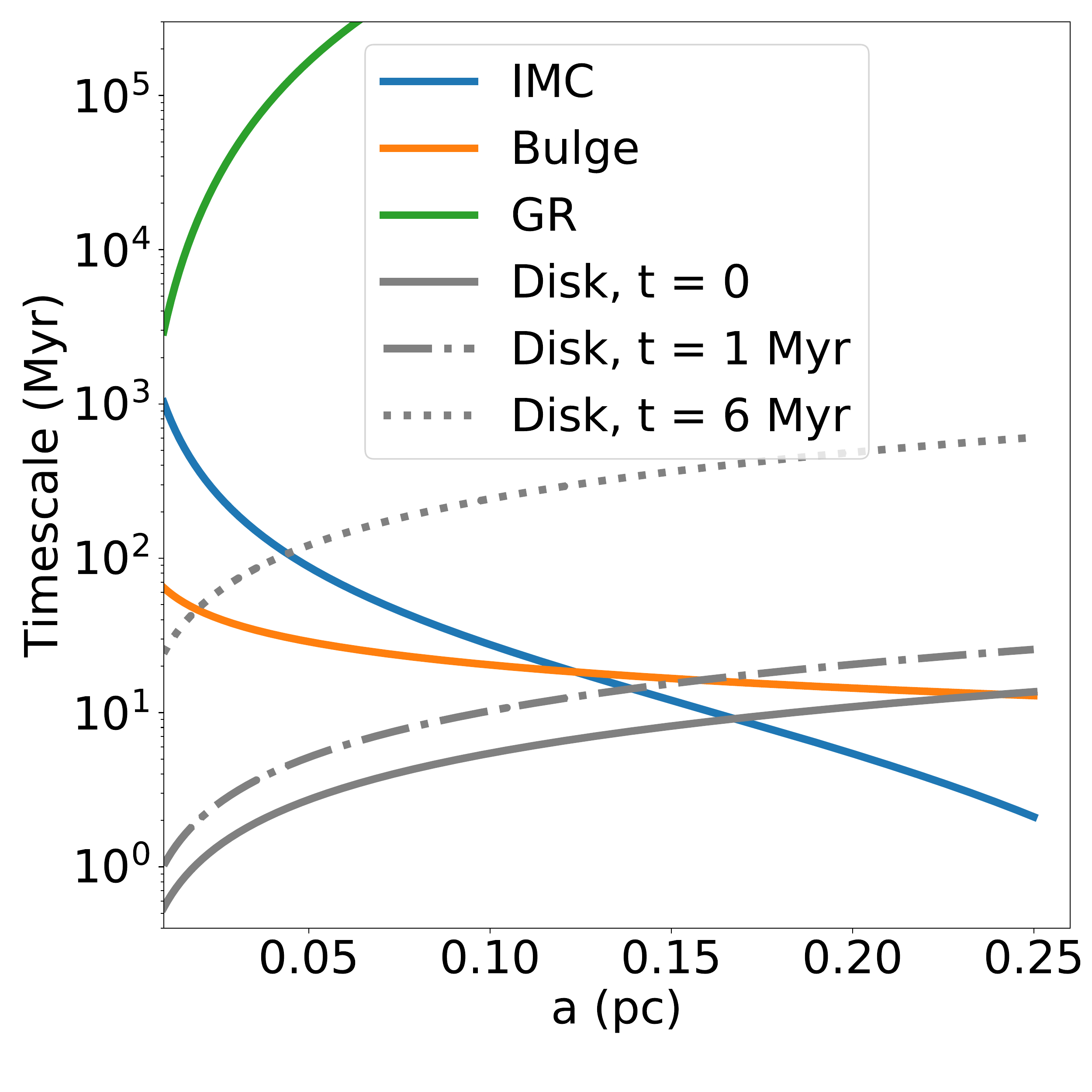}
\caption{The precession timescale of one disk star as a function of its initial semi-major axis. The precession timescale is computed with different components, including the GR effect (solid green line), the secular perturbation of an IMC which is initially located at 0.35 pc with eccentricity equal to 0.3 (solid blue line), and the influence of gaseous disk with $\Sigma_0=600$ g cm$^{-2}, \tau_{\rm dep}=1.58 $~Myr at various stages (grey lines).  
\label{fig:a_t}}
\end{figure}

We now add the contribution of the gas disk potential (with $\Sigma_0=600$ g cm$^{-2}$ as in Eq.~\ref{eq:B1B2}).  The value of $A^\prime_{\rm tot}$ is modified from those in Equation
(\ref{eq:atotbulge}) by subtracting a factor $\sim 2.2~ {\rm exp} (-{t / \tau_{\rm dep}})$ such that
\begin{equation}
 A^\prime _{\rm tot} \simeq -0.1
 \ \ \ \ \ \  {\rm or} \ \ \ \ \ \    A^\prime _{\rm tot} \simeq -2.1
\label{eq:atotdisk}
\end{equation}
at $t=0$ for co-orbiting and counter-orbiting IMC respectively.  Following the discussion in the previous section, 
these values imply that, at least initially, the co-orbiting IMC has the possibility of exciting large $e_\star$
than a counter-orbiting IMC (based on the magnitude of $e_{\star, limit}$ in Eq. \ref{eq:elimit} for these two cases).
However, after a few $\tau_{\rm dep}$, the gas disk's contribution diminishes whereas the bulge potential does not change (see Figure~\ref{fig:a_t}).
The values of $A^\prime _{\rm tot}$ reduce to those in Equation (\ref{eq:atotbulge}) so that the secular perturbation by 
the counter-orbiting IMC can lead to large asymptotic $e_\star$.  This adjustment is adiabatic in the limit $\tau_{\rm dep}
\gtrsim \tau_{\rm ecc}$  \citep{Nagasawa2003}.

The above estimate is made for one particular value of $a_\star$ (or equivalently $\alpha$).  In Equation (\ref{eq:atotprime}),
the contribution of the gas disk potential on the stars is proportional to $ \alpha^{-1} {\rm exp} (- t/\tau_{\rm dep}) $. During the
depletion of the gas disk, the significance of this term can be maintained at a smaller distance from the SMBH (see Figure~\ref{fig:a_t}).  For a
co-orbiting IMC, this inward propagation of the high-eccentricity-excitation region is referred to as the sweeping 
secular resonance.  Similarly, for a counter-orbiting IMC, the high-eccentricity excitation region sweeps outward.  
This property contrasts with the contribution from the Galactic bulge's potential which does not evolve with time.  

Figure \ref{fig:scenario} is a schematic illustration of the physical process associated with the excitation of the disk stars 
due to the sweeping secular resonance of a co-orbiting IMC. Within a few pc from \sgra, SMBH's point-mass gravitational 
potential dominates the 
stellar orbits.  Due to the IMC's secular perturbation, the stars' longitude of periastron liberates with eccentricity modulation
as the angular momentum is exchanged between the IMC's and stellar orbits.  The liberation period and amplitude are function of 
$a_{\rm IMC}$, $e_{\rm IMC}$, and $a_\star$  (Eq.~\ref{eq:dimensionless}).  Although the contribution from the Galactic disk and halo also 
leads to the precession of the stellar orbits, their effects are negligible within 1 pc from the SMBH.  In contrast, the gravitational 
perturbation from the Galactic bulge and/or a compact geometrically thin gaseous disk leads to a pronounced precession 
of the longitude of periastron 
for both the IMC and the stars with eccentric orbits. With modest gravitationally-stable $\Sigma$ distribution, 
the IMC and stellar precession rate are comparable to the liberation frequency it induces
on the nearby stars.  At some $a_\star$'s where these two frequencies resonate with each other, a finite angle of 
misalignment $\eta$ is maintained between the longitudes of periastron of the IMC and the stars. This asymmetry leads to angular
momentum transfer from the stars to the IMC and the excitation of $e_\star$.  

\begin{figure*}[ht!]
\epsscale{1}
\plotone{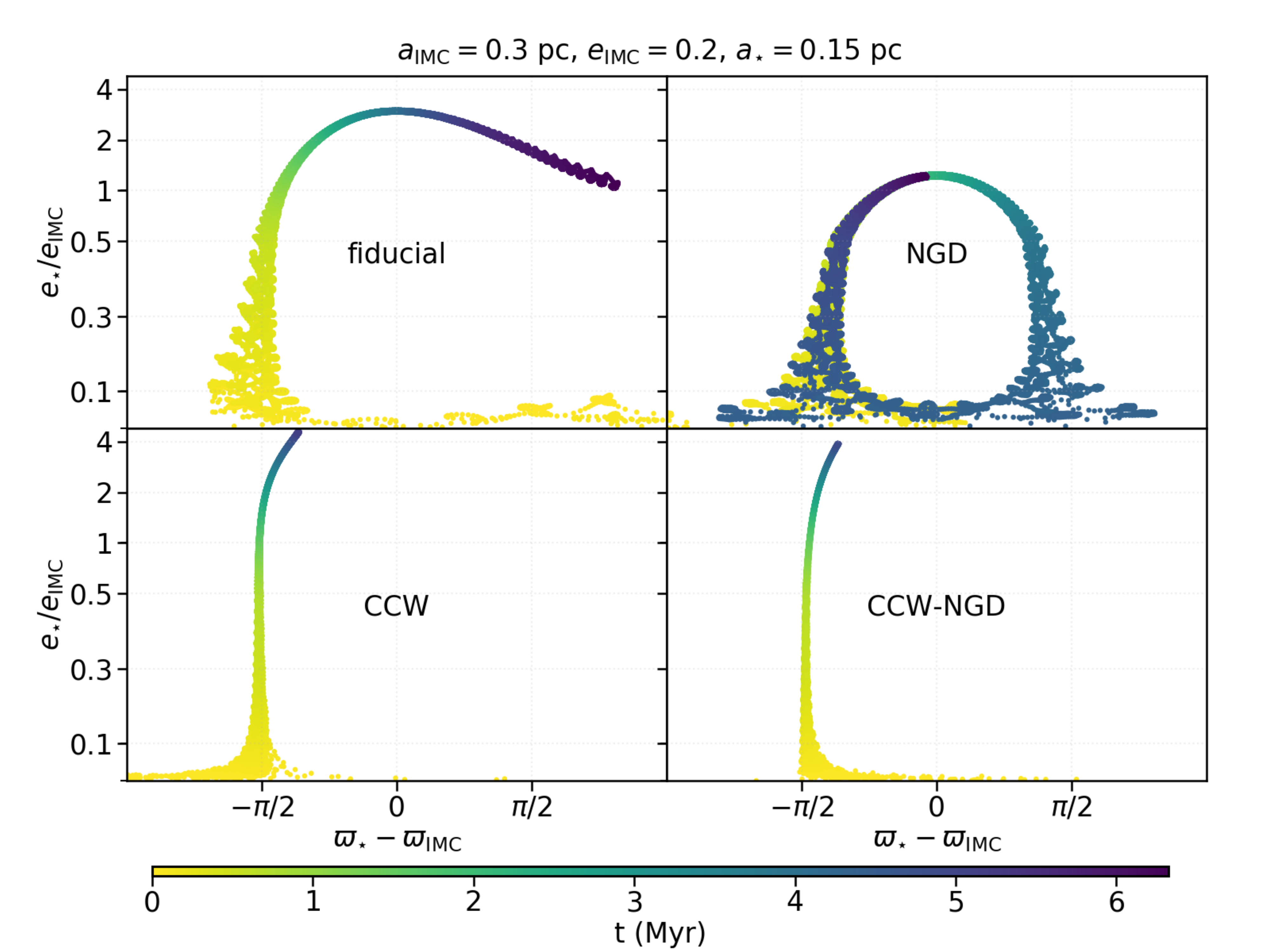}
\caption{The representative orbital evolution track in the ($\xi$, $\eta$)-diagram. We trace the disk stars in the fiducial model, NGD model, CCW model, and CCW-NGD model. In the top panels, two models represent a co-orbiting IMC; In the bottom panels, they show a counter-orbiting IMC. In all models, these disk stars were initially on a circular orbit at 0.15 pc, while the IMC is orbiting with semi-major axes around 0.3 pc and eccentricity equal to 0.2. The color labels the time evolution, we trace from 0 Myr to 6.3 Myr. 
\label{fig:pomega_e}}
\end{figure*}

The IMC's disk-induced precession frequency decreases during the depletion of the disk gas.  The location of the secular
resonance also propagates away from $a_{\rm IMC}$.  While the rate of eccentricity excitation does not depend on 
the disk depletion timescale, the asymptotic magnitude of $e_\star$ does, in the limit $\tau_{\rm dep} < 
\tau_{\rm ecc}$.  After the disk gas is severely depleted, the secular resonance sweeps
through an extended region of the disk.  We suggest this process is an efficient and robust eccentricity-excitation 
mechanism for a large fraction of stars that emerged out of the gaseous disk.   

\subsection{IMC with inclined orbit, Galactic disk, and halo}
In the analysis in \S\ref{sec:nogas} and \S\ref{sec:SSR}, we neglect contributions from the Galactic disk and halo
because their contribution to the precession of IMC and stars in the inner parsec is negligibly small compared with 
those due to the IMC, Galactic bulge, and gaseous disk.  In our numerical simulations, the validity of this approximation
is verified.  We also leave out, in the analytic study, the general possibility of an IMC with an inclined orbit relative
to that of the stellar disk.  A full 3D analysis introduces some additional degrees of freedom and dependent variables,  
including the inclination $i$ and longitude of ascending node $\Omega$.  Although it is straightforward to follow the approach
in celestial mechanics \citep{murray2000}, it requires the presentation of lengthy algebra.  For presentation purposes,
we reserve the treatment of inclined orbits to our numerical simulations in the next section.  


\section{Eccentricity excitation by sweeping secular resonance} \label{sec:model}

\begin{figure*}
\gridline{\fig{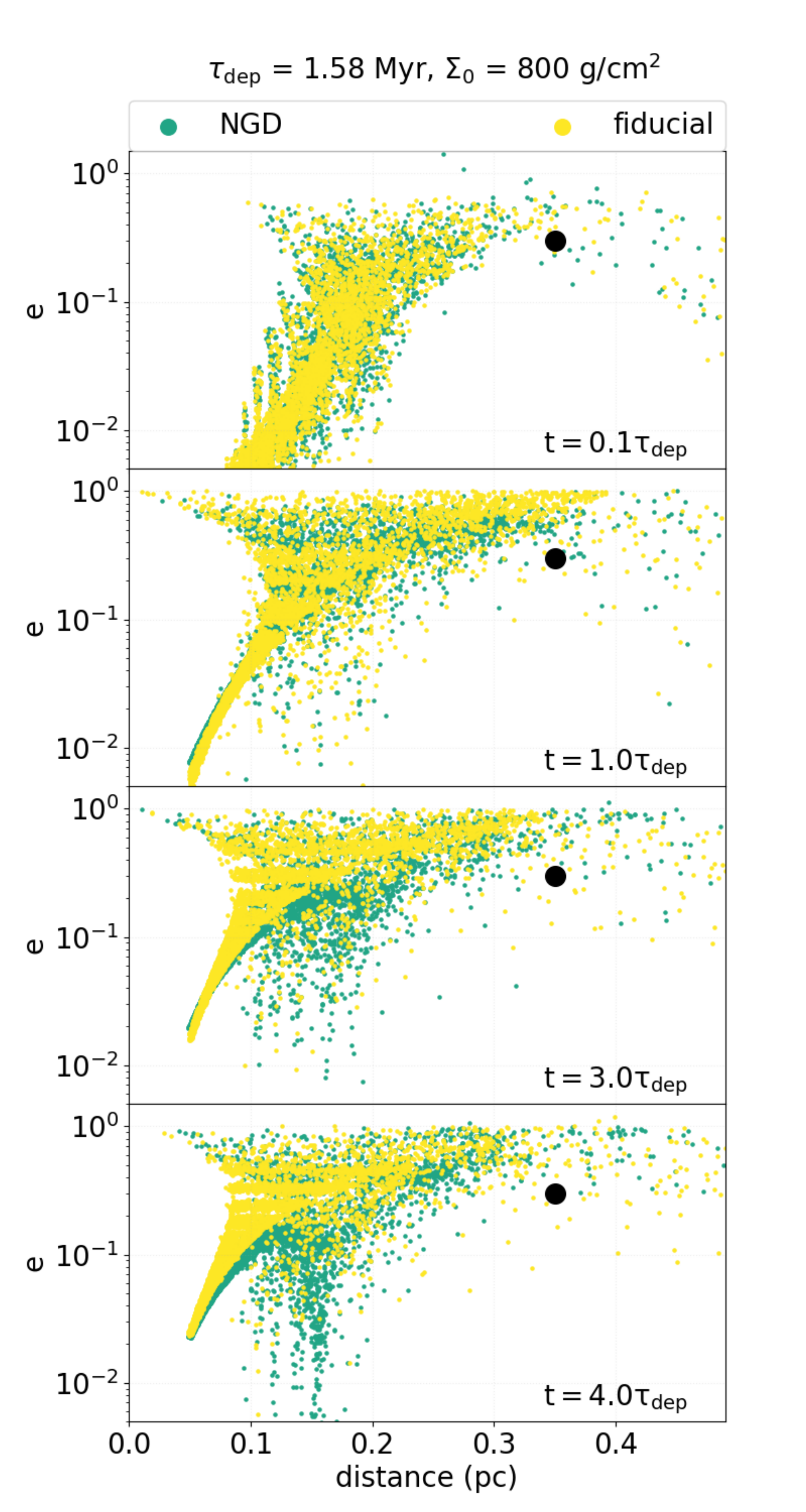}{0.5\textwidth}{(a)}
          \fig{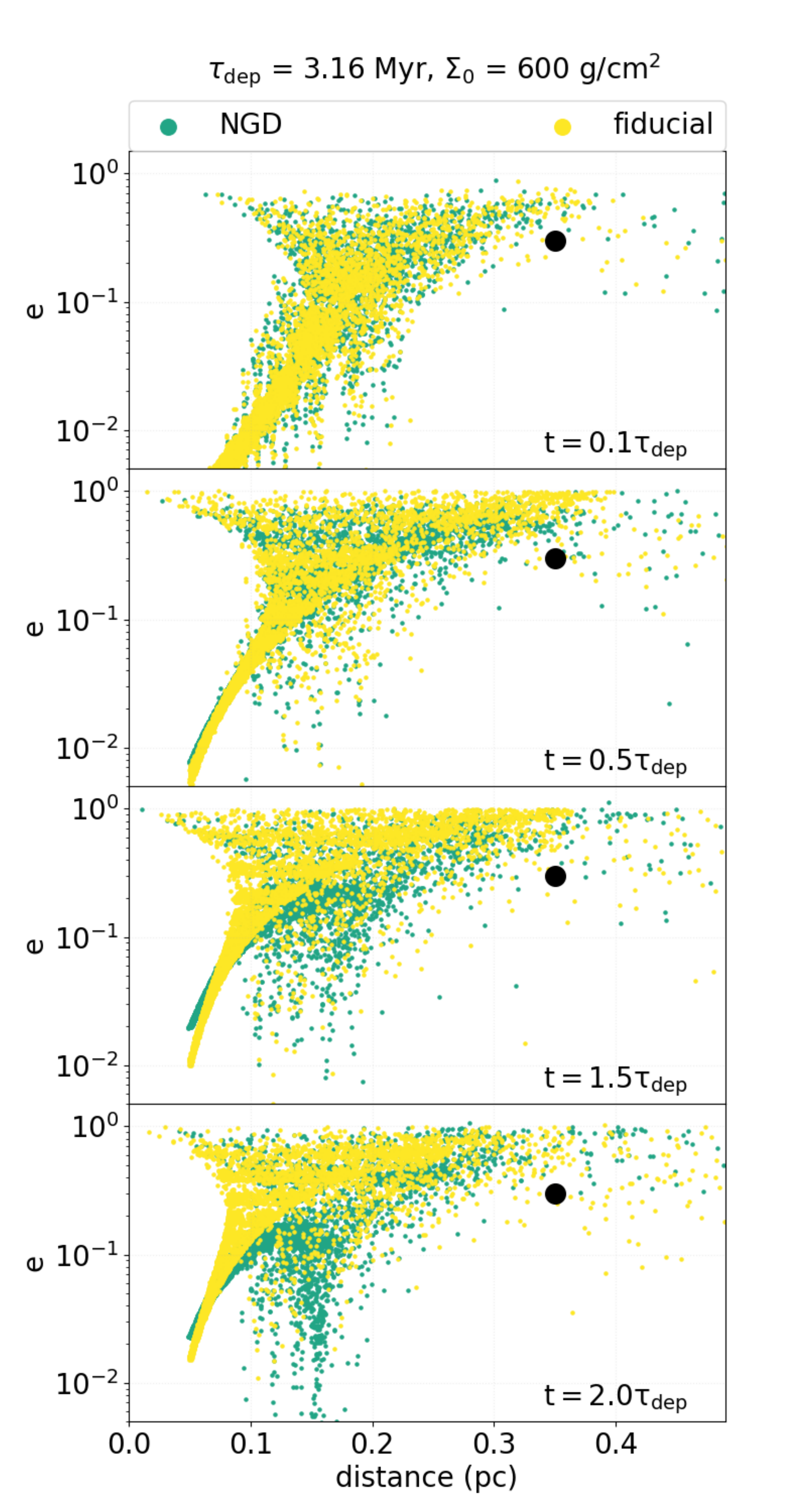}{0.5\textwidth}{(b)}
          }
\caption{The effect of the sweeping secular resonance on the eccentricity excitation of disk stars. The simulated results of the fiducial model and NGD model are represented with yellow and green dots, respectively. The black dot is a potential IMC located at 0.35 pc distance to \sgra~ with an eccentricity of 0.3. Disk stars are uniformed distributed around $0.05-0.5$ pc, except the region $a_{\rm in} - a_{\rm out}$. The left panels (a) represent the fiducial model with gas disk $\tau_{\rm dep} = 1.58$~Myr and $\Sigma_0 = 800~{\rm g/cm}^2$, the right panels (b) represent the result of gas disk with $\tau_{\rm dep} = 3.16$~Myr and $\Sigma_0 = 600~{\rm g/cm}^2$. 
\label{fig:as_es_t_com}}
\end{figure*}

The analysis in the previous section highlights the dominant effects of secular perturbation
and sweeping secular resonance.  We adopt several approximations to simplify the analytic treatment in the
limit of small to modest $e_\ast$.   The full effects can be calculated through the numerical integration 
of the Newtonian equation of motion with all or partial components of the contributing potentials, including 
the relativistic correction.  Based on these numerical simulations, we show here
how the IMC's secular perturbation may lead to the large eccentricity of CWSs.  In
paper II, we show that this process may also lead to the ejection of some hyper-velocity stars (HVSs).

\subsection{IMC's secular perturbation on a typical star}
\label{sec:numericalsec}

We first illustrate, with models NGD and CCW-NGD, the influence of IMC's secular perturbation on the disk stars in the 
absence of a gaseous disk.  These models are identical to cases we have analyzed analytically in \S\ref{sec:nogas}.
With $a_{\rm IMC} = 0.35$~pc and $e_{\rm IMC} = 0.3$, the IMC gravitationally scatter many nearby stars.  Their eccentricities 
quickly increase due to close encounters.  Over time, the eccentricities of more distant stars are also excited by IMC's secular 
perturbation.  In the previous section, we attribute this physical effect as the cause of CWSs' high eccentricity.

In order to compare with our analytic results (in \S\ref{sec:nogas} and \S\ref{sec:SSR}), we follow the evolution 
of a representative star which had an initial $a_\ast=0.15$~pc and $e_\ast=0$.  In the absence of a gaseous
disk, the evolution of $\eta (\equiv \varpi_\ast - \varpi_{\rm IMC}) $ and $\xi (\equiv e_\ast/e_{\rm IMC})$ in 
models NGD and CCW-NGD (for a co and counter orbiting IMC respectively) are plotted in the right panels of Figure 
\ref{fig:pomega_e}.  The discussion below Equation \ref{eq:etacrit} is verified by the results of model NGD which show 
the evolution of $\eta$ towards a critical value $\eta_{\rm crit}$ which is slightly larger than $-\pi/2$ in the limit of small $e_\ast$.  Also in agreement with Equation~\ref{eq:elimit}, $e_\ast$ makes
a transition from monotonically growth to liberation over a limited range of $\eta$ in model NGD.  The range of $\xi$ 
oscillates around limiting value $(e_{\ast, {\rm limit}} \sim 0.1$ or equivalently $\xi_{\rm limit} \sim 0.5$) on a time 
scale comparable to $\tau_{\rm ecc}$ (Equation~\ref{eq:tauecc}).  

The results in model CCW-NGD show a slower monotonic growth of $e_\ast$ (equivalent $\xi$).  The difference 
in the eccentricity excitation time scale between models NGD and CCW-NGD is due to
the difference in their magnitude of $A^\prime_{\rm tot}$ (Eq. \ref{eq:atotbulge}).
As we have indicated in \S\ref{sec:nogas}, model CCW-NGD is expected to produce a larger $e_{\ast, {\rm limit}}$ than
model NGD. If the calculation has continued beyond 6.3 Myr, this star would be ejected as its $e_\ast$ becomes
larger than unity.  

\subsection{Sweeping secular resonance due to gas depletion}
The fiducial and CCW models (left panels of Figure~\ref{fig:pomega_e})
for a co and counter orbiting IMC embedded in an evolving gaseous disk with 
the stars are equivalent to the cases we have analyzed analytically in \S\ref{sec:SSR}. 
The above discussion suggests that the gaseous disk's
contribution to the gravitational field modifies the precession rate of both IMC and the stars.  In both models,
the initial value of $A^\prime_{\rm tot}<0$ (Eq.~\ref{eq:atotdisk}) such that their 
$\eta_{\rm crit}$ is slightly smaller than $-\pi/2$ in the limit of small 
$e_\ast$ (Eq.~\ref{eq:etacrit}). During the depletion of the disk, $A^\prime_{\rm tot}$ increases to 
positive values so that $\eta_{\rm crit}$ becomes larger than $-\pi/2$ in both models.
In the fiducial model,
the star makes a transition from monotonic growth to circulation in which $\eta$ varies between $-\pi$ and
$\pi$.  The value of $e_\ast$ oscillates around a medium value which is larger than that in model NGD.
Had the simulation continued beyond 6.3 Myr, the circulation cycle would be completed.  

For the counter-orbiting IMC (model CCW), the initial surface density of the disk produces more negative values
of $A^\prime_{\rm tot}$.  Consequently the evolution of $\eta_{\rm crit}$ across $-\pi/2$ is delayed.  At the
end of the simulation, the propagation of the secular resonance promotes $e_\ast$ to attain a larger value
in model CCW than in model CCW-NGD.  The magnitude of $e_\ast$ continues to increase.  Had the simulation
continued beyond 6.3 Myr, this star would escape from SMBH's gravitational confine.

\subsection{Statistical evolution of disk stars}

The simulation of one representative star in the previous subsection is generalized to stars with 
a range of initial $a_\ast$.   Although their eccentricity is excited, most of the 
disk stars with $a_\star \lesssim 0.2$~pc in the NGD model, have asymptotic $e_\star \lesssim 0.1$
(green dots in Figure~\ref{fig:as_es_t_com}), consistent with our estimate for 
$e_{\rm \star, limit}$ for the co-orbiting IMC (Eq.~\ref{eq:elimit}). These stars'
modest liberation amplitudes limit their asymptotic $e_\ast$ to be smaller than that observed. 


The effect of sweeping secular resonance is noticeable in the fiducial model (yellow dots in Figure~\ref{fig:as_es_t_com}).
At the end of the simulation, stars in the fiducial model generally attain larger $e_\ast$ than those in the NGD model.  As we have indicated in \S\ref{sec:SSR}, for a co-orbiting IMC, sweeping secular resonance
is required to excite the eccentricity of disk stars to the observed values.

\begin{figure*}
\gridline{\fig{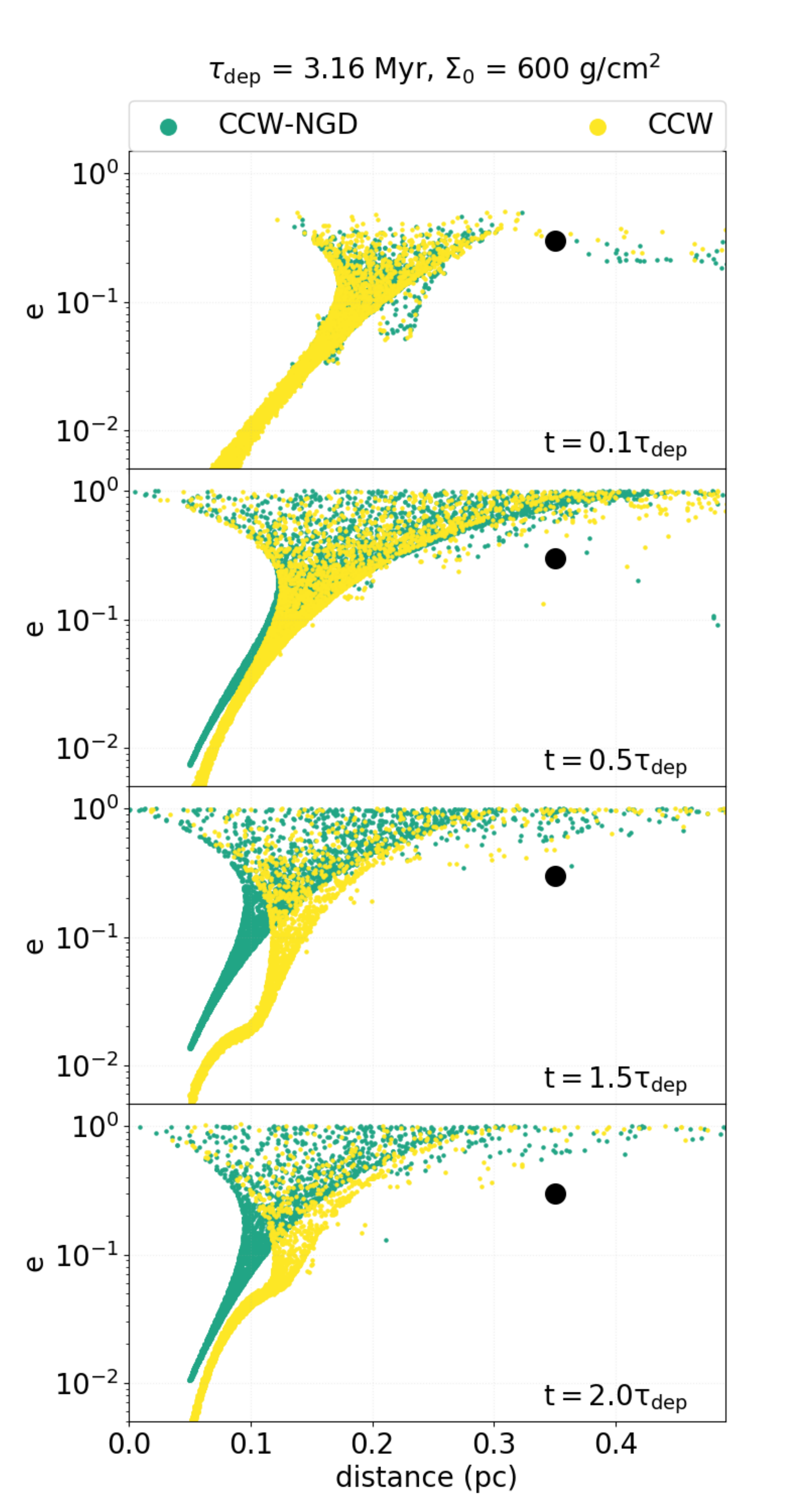}{0.5\textwidth}{(a)}
          \fig{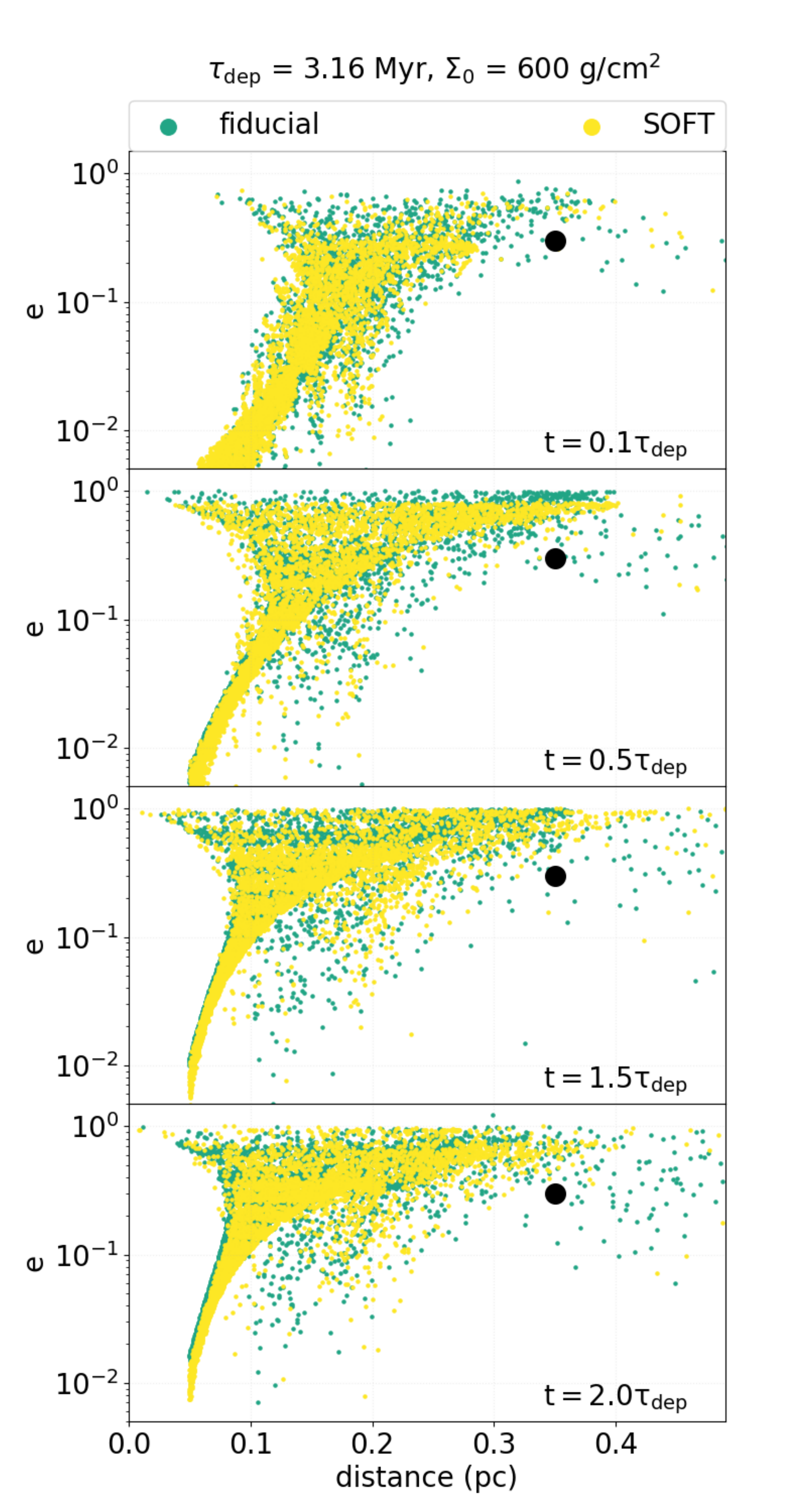}{0.5\textwidth}{(b)}
          }
\caption{Same as Figure~\ref{fig:as_es_t_com}, but compare the fiducial model, SOFT model, CCW model, and CCW-NGD model. The left panels (a) represent the CCW model and CCW-NGD model. The right panels (b) represent the result of the fiducial model and SOFT model. 
\label{fig:as_es_t_com1}}
\end{figure*}

For a counter-rotating IMC, the presence of the disk slows down the propagation of the secular resonance.
The delayed eccentricity excitation in model CCW (with gaseous disk) to CCW-NGD (without gaseous disk) 
is contrasted by the yellow versus green dots in the left panel of Figure~\ref{fig:as_es_t_com1}.  At $t > > \tau_{\rm dep}$,
the precession of IMC and stars in model CCW reduces to that in the CCW-NGD model.  For the counter-orbiting
IMC, stars acquire large eccentricities with or without the gaseous disk.  

We also compare the SOFT and the fiducial model in the right panel of Figure~\ref{fig:as_es_t_com1}, and find that whether the IMC is a point-mass like IMBH or a compact stellar cluster, it can produce comparable orbital excitation for these residual disk stars. However, an equal mass IMBH and a softening stellar cluster have the different capacity in stimulating high-velocity escapees, like HVSs. We intend to discuss related efficiency in paper II of this series.

\subsection{Gaseous disk's mass and depletion timescale}

\begin{figure*}[ht!]
\epsscale{1.1}
\plotone{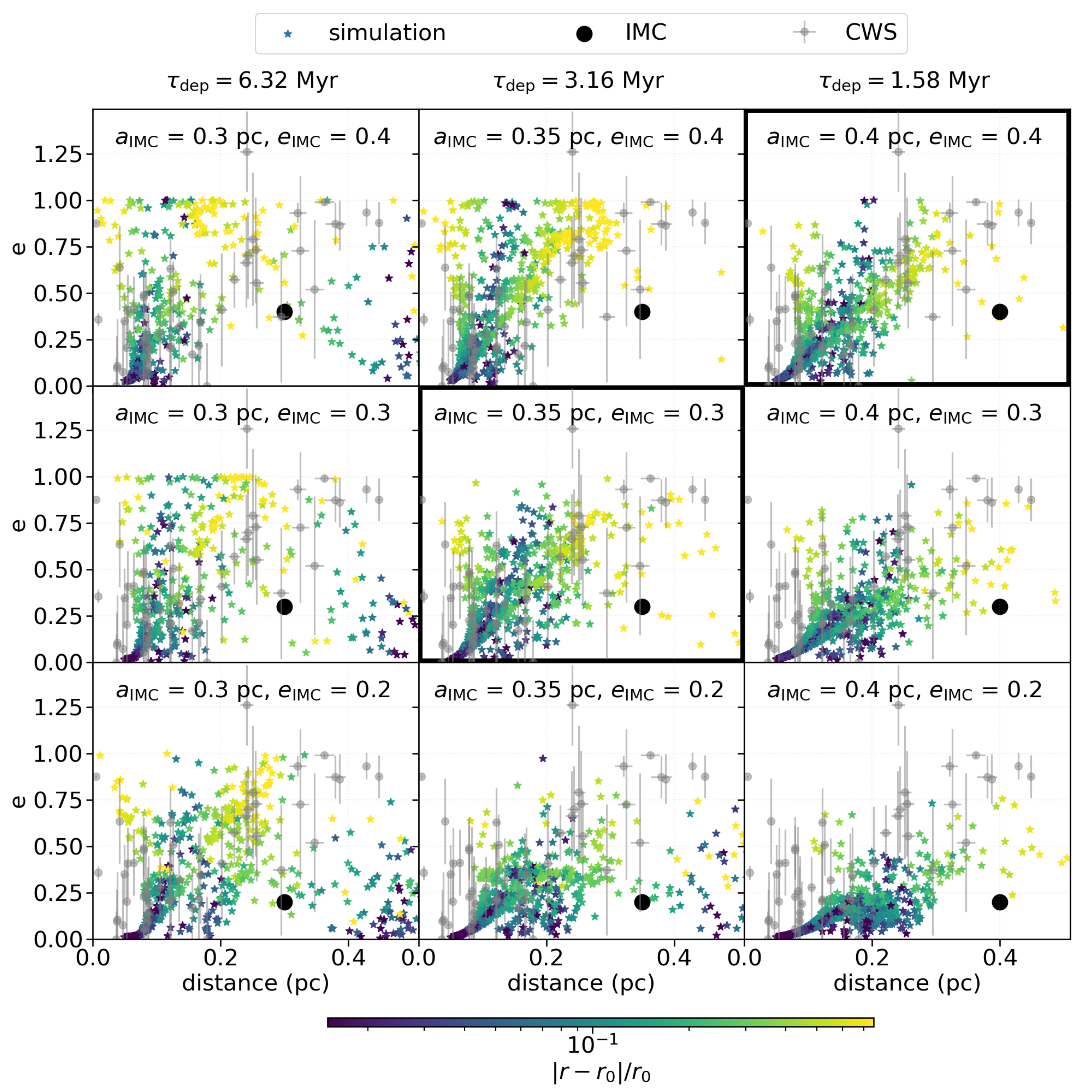}
\caption{Eccentricity of clockwise disk stars as a function of their distance from the SMBH. The filled grey dots with error bars represent the observed CWSs as given by \cite{paumard2006}. The black dot depicts the possible orbital parameters ($a_{\rm IMC}, 
e_{\rm IMC}$) of an IMC.  In the left, middle, and right columns, we adopt $a_{\rm IMC}=0.3$ pc with $\tau_{\rm dep}= 6.32~$Myr,
$a_{\rm IMC}=0.35$ pc with $\tau_{\rm dep}= 3.16~$Myr, and $a_{\rm IMC}=0.4$ pc with $\tau_{\rm dep}= 1.58$~Myr, respectively.
In the top, middle, and bottom rows, we choose $e_{\rm IMC} = 0.4, 0.3, ~{\rm and}~0.2$ respectively.  The filled stars show 
the $e_\star$ and $a_\star$ of the disk stars at $6$~Myr. Colors label the fractional change in the disk stars' distance 
from the SMBH.
\label{fig:aj_ej_com}}
\end{figure*}

For comparison, we simulated two versions of the fiducial model with the same $M_{\rm IMC}$, $a_{\rm IMC}$, and $e_{\rm IMC}$.  
In these two models, the gaseous disks are assumed to have a) $\Sigma_0=600$ g cm$^{-2}$ and $\tau_{\rm dep} = 3.16$ Myr and b)
$\Sigma_0=800$ g cm$^{-2}$ and $\tau_{\rm dep} = 1.58$ Myr (see yellow dots in the right and left  panels, respectively, of 
Figure~\ref{fig:as_es_t_com}).  The results of both models show that, within a few $\tau_{\rm dep}$, $e_\star$ is excited 
to $\gtrsim 0.3$ for most of the stars with 0.1~pc $\leq a_\star \leq$ 0.4~pc.  These values are comparable to those observed
(see \S\ref{sec:constraint}).  These large eccentricities not only exceed those
in model NCD, but they are extended over much wider radial distances from the SMBH.  This outcome is due to the extensive
region swept by IMC's secular resonance during the depletion of the gas disk, as shown quantitatively in Equations~(\ref{eq:dimensionless}) and (\ref{eq:atotprime}) and quantitatively discussed in \S~\ref{sec:SSR}.

For illustration purposes, we assume a hypothetical $\Sigma$ distribution in our numerical models.  The two 
chosen values of $\Sigma_0$  provide two different starting locations for IMC's secular resonance 
(Eq.~\ref{eq:pratio}).  The magnitude of $\tau_{\rm dep}$ determines the sweeping speed of IMC's secular resonances (Eq.~\ref{eq:atotprime}).  Protracted sweep (with larger $\tau_{\rm dep}$) prolongs the duration of angular momentum exchange and enhances the magnitude of excited eccentricity.  Similar asymptotic values and distribution of $e_\star$ are excited by a range of
combined $\Sigma_0$ and $\tau_{\rm dep}$ values.  Due to this degeneracy, we cannot constrain from the observed $e_\star$ distribution,
the gas disk structural, and evolutionary parameters to some unique values.  Nevertheless, these results also show that eccentricity excitation by sweeping secular resonance is generic and robust. 
  
\subsection{IMC's eccentricity and semi-major axis}
\label{sec:constraint}

The observed $e_\star$ distribution can be used to provide some constraints on $a_{\rm IMC}$ and $e_{\rm IMC}$.
In Figure~\ref{fig:aj_ej_com}, we compare the orbital distribution of 1000 disk stars under the influence of an 
co-orbiting IMC's 
secular perturbation. In the fiducial models, we choose for IMC a mass $M_{\rm IMC}=10^4 M_\odot$ with various values of 
$a_{\rm IMC} (=0.3, 0.35, 0.4$ pc) and $e_{\rm IMC} (= 0.2, 0.3, 0.4)$ in the same orbital plane as the gaseous disk and 
the stars' initial orbital plane.  

The disk stars are set initially with zero $e_\star$ and random $a_\star$ distribution between $0.05-0.5$~pc uniformly.  For the 
purpose of highlighting the influence of sweeping secular resonance, stars initially in the close-encounter-with-the-IMC 
zone (between $a_{\rm in} - a_{\rm out}$) are excluded.

The observed $a_\star$ and $e_\star$ are plotted as grey dots in Figure \ref{fig:aj_ej_com}.  Up to ($\sim 80 \%$) CWSs 
stars are located between $0.05-0.5$~pc and the high $e_\star$ ($>0.5$) stars are mostly located between 0.2~pc and 0.5~pc.  Unless IMC has $e_{\rm IMC} \sim 0.2-0.4$ and $a_{\rm IMC} \sim 0.3-0.4$ pc, it 
is difficult to excite the eccentricity of disk stars to such high eccentricities.   
These numerical results are consistent with our analytic calculation in \S~\ref{sec:SSR}.  

All simulations are carried out over $\sim 6$~Myr with $\Sigma_0 = 600$ g cm$^{-2}$.  In order for IMC's secular resonance to sweep through most of regions with $a_\star =0.05-0.5$~pc for all the cases, we choose $\tau_{\rm dep} = 6.32, 3.16,$ and 1.58 Myr for $a_{\rm IMC} 
= 0.3, 0.35$ and $0.4$ pc respectively.
 
The results of these parameter study show two sets of preferred models which match the observational data:
1) $a_{\rm IMC} = 0.35$~pc with $e_{\rm IMC} = 0.3$, and 2) $a_{\rm IMC} = 0.4$~pc with $e_{\rm IMC} = 0.4$
(see Figure~\ref{fig:aj_ej_com}).  In models with small $a_{\rm IMC} (\leq 0.3$~pc), the eccentricity 
excitation for $a_\star \geq a_{\rm IMC}$ is below ($\leq 0.2-0.5)$ the observed $e_\star$ especially
for the disk stars around $ 0.4-0.5$ pc. Similarly, in models with large $a_{\rm IMC} 
(\geq 0.4$~pc) and small $e_{\rm IMC} (\leq 0.3)$, the eccentricity excitation for $a_\star < a_{\rm IMC}$ 
is mostly below the observed $e_\star$ especially for the disk stars around $0.1-0.2$~pc.
(The probability of observing an IBH or dense stellar cluster with $a_{\rm IMC} > 0.4$ pc and
$e_{\rm IMC} >0.4$ at the current projected separation of 0.13~pc from the \sgra~ is relatively low).
For $a_{\rm IMC} \sim 0.35-0.4$~pc, modest $e_{\rm IMC} (\sim 0.3-0.4)$, IMC is able to excite the eccentricity
of the disk stars with an extended $a_\star$ ($\sim 0.05-0.5$ pc) to the observed values after a few Myr.

Similar constraints can also be obtained for a counter orbiting IMC.  Since the initial gravitational
potential in model CCW-NGD (and the asymptotic configuration in model CCW) already produce
small $A^\prime_{\rm tot}$ and large $e_{\ast, {\rm limit}}$ (Eqs. \ref{eq:atotbulge} and \ref{eq:elimit}),
a more extended range of suitable $a_{\rm IMC}$ and $e_{\rm IMC}$ can lead to large asymptotic $e_\ast$.

\begin{figure*}[ht!]
\epsscale{1.}
\plotone{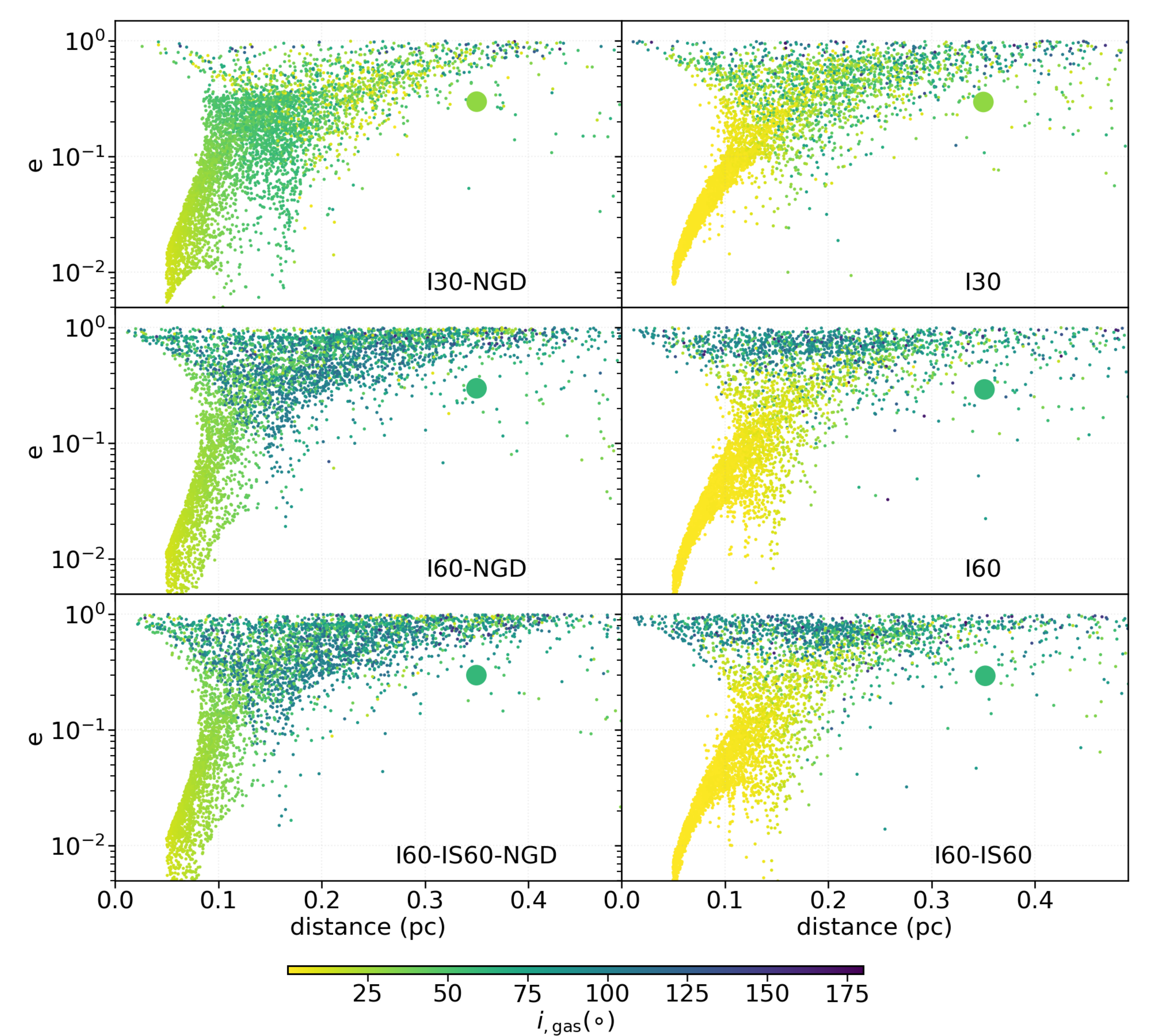}
\caption{The eccentricity distribution of simulated stars around 6 Myr as a function of distance to the GC in the I30, I30-NGD, I60, I60-NGD, I60-IS60 and I60-IS60-NGD models. Colors label the inclination of disk stars (and IMC). $i_{,\rm gas}$ refers to the mutual inclination between the orbits of the disk star (or IMC) and the gas disk. For these NGD models, it refers to the inclination of the disk stars (or IMC) relative to the disk stars' initial orbital plane. The dot size labels the mass of disk stars and IMC.
\label{fig:d_e_inc}}
\end{figure*}

\subsection{IMC with an inclined orbit}
We also carry out two series of simulations with an inclined IMC orbit, relative to the initial stars' (or gaseous disk)
orbital plane by $i_0=30^{\circ}$ and $60^{\circ}$.  In celestial mechanics, secular interaction between planets with inclined
orbits leads to modulations in their inclination and eccentricity which are correlated through the approximate conservation of 
the Delaunay momentum $H_K = {\sqrt {1-e^2}} ~{\rm cos~} i$ \citep{murray2000}. 
The inclination in this expression refers to that relative to an invariant plane. This conservative relationship 
is also applicable to analogous stellar systems \citep{Naoz2016}.
With sufficiently large mutual inclination, $i$, 
it is possible for the precession frequencies of periastron and ascending node to enter into a Lidov-Kozai resonance 
such that nearly parabolic eccentricity can be excited with a corresponding change in $i$ \citep{Lidov1962,  Kozai1962, 
Innanen1997}.  
Accordingly, even if starting with a low-inclination configuration, the orbit of an inner disk star can still be excited to high values.  If stars which undergo close encounters with the SMBH, it is possible for their orbits to flip from 
prograde to retrograde orientation due to the chaotic perturbation of such an eccentric IMC, as discussed as in 
the context of the eccentric Kozai-Lidov mechanism \citep{Li2014a, Li2014b, Naoz2016, Hansen2020}.
With the contribution of the Galactic bulge and the gaseous disk to the potential, the precession rates
in the present context are modified from those associated with classical celestial mechanics.  It also enlarges the
amplitude of variation of the Delaunay momentum during a secular cycle.

In the top two rows of Figure (\ref{fig:d_e_inc}), we plot the instantaneous radial distribution of $e_\ast$ at the end 
of the simulation for models I30-NGD, I30, I60-NDG, and I60. 
Although the results in model NGD show that $e_\ast$ is excited with $i_0=0$, a comparison between models NGD, I30-NGD, and 
I60-NGD indicates that a larger $i_0$ can enhance the oscillation amplitude in both $e_\ast$ and $i_\ast$.  However, 
a fraction of stars attains nearly parabolic $e_\ast$ with large changes in $i_\ast$ from the weakly perturbed stars.

In models I30 and I60 with gaseous disks, we assume the IMC cannot open a gap.  Consequently, its precession due to
the disk potential is in the same direction as the stars.  The longitudinal precession rates of both periapse 
and ascending node evolve during the depletion of the disk.  The amplitudes of $e_\ast$ and $i_\ast$ modulation 
are enhanced for stars with $a_\ast \gtrsim 0.2$~pc and suppressed for those with $a_\ast \lesssim 0.13$~pc, especially 
in model I60. The frequency of stars with highly eccentric and inclined orbits is increased by the perturbation of 
a highly inclined IMC.  This diversity may account for the dispersion in the inclination of the young stars around
\sgra.

In all the models we have considered so far, we assume the angular momentum vector of the gaseous disk (and that of 
the stars initially) is parallel to that of the Galactic disk.  In the last row of Figure (\ref{fig:d_e_inc}), we introduce model I60-IS60-NGD and I60-IS60.  In these models, the stellar disk is initially $60^o$ inclined
to the Galactic disk (Table \ref{tab:models}).  Since the scale length for the Galactic disk is much larger
than $a_\ast$, we do not expect any significant difference between models I60-IS60-NGD and I60-NGD (as well
as between models I60-IS60 and I60) as shown in  Figure (\ref{fig:d_e_inc}).  This agreement justifies the 
neglect of the Galactic disk and halo in the analytic calculation in \S\ref{sec:planetaryeq}.

\section{Summary \& Discussion}
\label{sec:summary}

There is a population of very young and massive stars within the central 0.5 pc of the Galaxy.
Many of them orbit clockwise around the SMBH on the same plane.  These CWS's, especially
those with semi-major axis $\sim 0.2 - 0.5$ pc, are observed to have large 
eccentricities relative to the SMBH ($e_\star \gtrsim 0.5$, see Figure~\ref{fig:aj_ej_com}).   
If they were formed in a nearly Keplerian disk with modest or negligible eccentricities, there 
would not be adequate time for the eccentricity of these massive stars to be excited, by their
intrinsic dynamical relaxation process \citep{lightman1977, rauch1996, merritt2013}, to the 
observed values within their expected life span.  

In this paper, we mainly focus on the possibility of a companion to the SMBH as the culprit 
of eccentricity excitation for the CWSs.  We suggest a potential candidate, IRS 13E, which
has a projected separation of 0.13 pc from \sgra.  It has an estimated mass $\sim 10^4 M_\odot$
and somewhat uncertain orbital motion around the SMBH in the counter-clock direction.  There 
have been suggestions that it may be an IMBH or a compact stellar cluster.  We refer this 
satellite perturber as an intermediate-mass companion, IMC.  
Since we are interested in its secular perturbation 
over a wide range of distances (a fraction of a pc), our analysis applies regardless whether 
it is an IMBH or a compact stellar cluster with a comparable mass.  

We consider a range of semi-major axis ($a_{\rm IMC} = 0.3-0.4$ pc) and eccentricity 
($e_{\rm IMC} = 0.2-0.4$) for the IMC.  
In order to circumvent the large observational uncertainty in its orbit, we consider 
both co-orbiting and counter-orbiting IMC which revolves around the SMBH either in 
the same or opposite direction as the CWSs.  In paper II, we also examine the possibility of 
an IMC with inclined orbits relative to the CWSs. 

We show there are several stages of stars' dynamical evolution.  We assume the CWSs emerged 
in their natal disk with negligible initial eccentricity. Under the IMC's secular perturbation,
their relative longitude of periastron $\eta$ (with respect to that of the IMC) rapidly evolve to some quasi finite (non zero) values, initially slightly larger 
than $-\pi/2$.  In this secular resonant state, the CWSs' precession is nearly 
synchronized with that of the IMC.  Under the IMC's perturbation, CWSs'
eccentricity $e_\star$ increases monotonically as their $\eta$ increases toward $0$
with a secular growth rate $\sim e_{\rm IMC} n_{\rm IMC} M_{\rm IMC} / (4 M_{\rm SMBH})$.
When $e_\star$ reaches an asymptotic limit $e_{\rm \star, limit}$, the stars exit their 
secular resonances with the IMC as the evolution of their $\eta$ transforms from 
a protracted shift of quasi-equilibrium values to liberation within a limited range 
or circulates over $2\pi$.  During these secular cycles, the stars' eccentricity 
is modulated by the exchange of angular momentum between them and the IMC.  If 
the potential of all the perturbing components remains constant, there would 
not be any additional systematic eccentricity evolution beyond $e_{\star, {\rm limit}}$
other than the modulation associated with liberation and circulation.

Our analysis shows that the eccentricity growth 
limit ($e_{\rm \star, limit}$) depends on the delicate balance between
IMC and CWSs' precession frequencies.  Both of these frequencies are partially 
determined by the IMC's and stars' relative direction of motion around the SMBH, 
the stationary mass ratios between the Galactic bulge, SMBH, and IMC.  
They also depend on the stars' semi-major axis ($a_\star$), the ratio of CWSs' and 
IMC's eccentricity ($\xi$), and the evolving mass of the gas disk ($M_D$ exp(-t
/$\tau_{\rm dep}$).  In the absence of an evolving gaseous disk, $\eta$ makes a
transition from the monotonic, slow shift of quasi-equilibrium to liberation or circulation 
with relative small $e_{\rm \star, limit}$ for a co-orbiting IMC.  But, under the 
secular perturbation of a counter-orbiting IMC, the transition occurs for many stars, 
in the $0.2-0.5$ pc region,  with much larger $e_{\rm \star, limit} (\gtrsim 0.5)$.

The youthfulness and the coplanar orbits of the CWSs suggest that they have emerged 
from a gaseous disk.   Here,
we assume an initial mass for the disk and its exponential 
depletion on a timescale of a few Myr.  We speculate that this disk may be the reservoir 
of the Fermi bubble without specifying how it was launched.  While the disk had a mass comparable to that of the IMC,  its 
contribution to the gravitational potential significantly modified the stars' and 
IMC's precession rates from those due to their mutual secular interaction
and the Galactic bulge potential.  In this limit, $e_\star$'s transition from 
monotonic growth to liberation or circulation occurs with relative large 
$e_{\rm \star, limit}$ for a co-orbiting IMC and with modest 
$e_{\rm \star, limit}$ for a counter-orbiting IMC.
 
Since there is no sign of a gaseous disk today, it must have depleted after the formation
of the CWSs. During the depletion of the gaseous disk, its influence on the IMC's and stars'
precession wanes.  This transformation extensively reduces/enlarges the 
high-$e_{\rm \star, limit}$ domain for a co-orbiting / counter-orbiting IMC respectively.  
We refer to this evolving landscape as the sweeping secular resonance. We carry out several 
series of systematic numerical simulations to demonstrate that the sweeping secular resonance 
can facilitate the potential excitation of $e_\star$ to the observed values ($\gtrsim 0.5-0.7$) 
of CWSs or the hyperbolic orbits of HVSs.  This sweeping secular resonance process is effective 
for both a co-orbiting and counter-orbiting IMC.  But, without a gaseous disk, there would
be no sweeping secular resonance, and the precession of a co-orbiting IMC alone 
would not ensure the excitation of adequately large $e_\star$.

Based on this scenario, we compare the results of the numerical simulation with the
observed eccentricity distribution of the CWSs to constrain the IMC's orbital
semi-major axis to be in the range of $0.3-0.4$~pc  which is somewhat larger than IRS 13E's projected
separation from \sgra as suggested by the X-ray morphology of its surrounding region
\citep{wang2020}.  We also find that for a co-orbiting IMC, the effectiveness 
of this process requires the depletion timescale ($\tau_{\rm dep}$) to be comparable 
to or larger than a fraction of the liberation timescale.  Since a counter-orbiting
IMC can excite high $e_{\rm \star, limit}$ without a disk, this condition is not 
required.  Our numerical simulations also demonstrate that since its secular perturbation
is applied through distant torque rather than short-range close encounters, the 
perturbing companion (i.e., IRS 13E) can be an IMBH or a compact bound stellar cluster
with the same mass.  

Our analysis highlights the sensitive dependence of CWSs' eccentricity on the IMC's 
mass and orbit.  Yet the physical nature and integrity of our suggested candidate, 
IRS 13E,  remain uncertain and controversial.
Future observational determination of these parameters will be useful
to constrain this model.  Other poorly known quantities are the mass
and depletion timescale of the hypothetical gaseous disk.  A complementary model
for the origin of the Fermi bubble may provide insights into the origin and 
evolution of the CWSs. In paper II of this series, we will also study the population of HVSs in great detail.

In the work presented here, we assume a stable orbit of an eccentric IMC throughout the work. However, the dynamical friction by the surrounding stars leads to an orbital decay of the IMC in central regions of the Galaxy, and the final decay inside the central parsec occurs within $\sim 15$ Myr \citep{Just2011}. Theoretically, the orbital migration of the IMC induces additional secular resonance sweeping through the central parsec, in analogy to the effect of a depleting gaseous disk.  This effect has been studied in the 
solar system context \citep{agnor2012}. Also, we do not include binary configurations in this work. However, the existence of a binary population may influence the kinematic measurements, as discussed in \cite{Naoz2018}. And the survival of any close binaries with hyper-velocity can provide a discriminate test between our model and the Hill's mechanism.
We will address these issues in the near future.

\acknowledgments
Simulations in this paper made use of the REBOUND code which is publicly available 
at \href{http://github.com/hannorein/rebound}{REBOUND} website.  We thank Tuan Do,
Stefan Gillessen, Mark Morris, and Hangci Du for useful conversation.  
This work is partly supported by the National Key Research and Development Program of China (No. 2018YFA0404501 to SM), by the National Science Foundation of China (Grant No. 11821303, 11761131004 and 11761141012 to SM) and by the China Postdoctoral Science Foundation (Grant No. 2017M610865 to XCZ).

\vspace{5mm}

\software{REBOUND \citep{reinliu2012, rein2019}}

\vspace{15mm}

\bibliography{zxc_ecc.bib}{}
\bibliographystyle{aasjournal}

\end{CJK*}
\end{document}